\documentclass[runningheads]{llncs}
\usepackage{graphicx}
\usepackage[colorlinks]{hyperref}
\usepackage{enumerate}
\usepackage{subcaption}
\usepackage{multirow}
\usepackage{subcaption}
\usepackage{tabularx}
\usepackage{amssymb}

\begin{document}
\title{Evaluation of Software Product Quality Metrics}
%
%
\author{Arthur-Jozsef Molnar \and Alexandra Neam\c{t}u \and Simona Motogna}
\institute{Faculty of Mathematics and Computer Science, Babe\c{s}-Bolyai University, Cluj-Napoca, Romania
\email{\{arthur, motogna\}@cs.ubbcluj.ro, nais1841@scs.ubbcluj.ro}}

\maketitle              
\begin{abstract}
Computing devices and associated software govern everyday life, and form the backbone of safety critical systems in banking, healthcare, automotive and other fields. Increasing system complexity, quickly evolving technologies and paradigm shifts have kept software quality research at the forefront. Standards such as ISO's 25010 express it in terms of sub-characteristics such as maintainability, reliability and security. A significant body of literature attempts to link these subcharacteristics with software metric values, with the end goal of creating a metric-based model of software product quality. However, research also identifies the most important existing barriers. Among them we mention the diversity of software application types, development platforms and languages. Additionally, unified definitions to make software metrics truly language-agnostic do not exist, and would be difficult to implement given programming language levels of variety. This is compounded by the fact that many existing studies do not detail their methodology and tooling, which precludes researchers from creating surveys to enable data analysis on a larger scale. In our paper, we propose a comprehensive study of metric values in the context of three complex, open-source applications. We align our methodology and tooling with that of existing research, and present it in detail in order to facilitate comparative evaluation. We study metric values during the entire 18-year development history of our target applications, in order to capture the longitudinal view that we found lacking in existing literature. We identify metric dependencies and check their consistency across applications and their versions. At each step, we carry out comparative evaluation with existing research and present our results.
\end{abstract}

\section{Introduction}
Software development has experienced an exponential increase over the past decades, which can be observed in the variety of applications available (such as web, mobile, real time and so on), as well as in application size and complexity. Large-scale and enterprise applications are being developed over longer periods of time, using larger teams that are in many cases geographically distributed. In the same time frame, project management and software methodologies, available tools and development environments have evolved in an attempt to keep the pace with increasing requirements.

This increase in size and complexity raises another problem, namely the necessity to control the software development processes, and implicitly to measure it, as "you cannot control that which you cannot measure" \cite{2}. In accordance, the domain of software metrics has evolved both as methodology as well as in terms of available software products, being influenced by the development of programming languages, paradigms and methodologies. 

Software quality assurance is also an important aspect as software products have to satisfy user needs related to ease of use, security and reliability. Furthermore, development related needs such as maintainability, portability and testability must also be accounted for. The latest software quality models have undergone standardization processes, such as ISO standards 9126 and 25010, in order to establish a set of common criteria for software products. These standards can significantly benefit from data provided by software metrics, as there exists consistent research results that report the influence of software metrics on software quality factors \cite{6,5,36,11,20}. 

However, additional data analysis is required before general models can be built \cite{24}. Also, even if the influence of metrics on quality factors is well understood and accepted, there does not yet exist any general accepted method to evaluate software quality factors based on software metric values. As such, the relation between metric values and software quality factors remains an open problem. We aim to address this issue in the present paper. We carry out a comprehensive evaluation on values of software metrics that are widely associated with software product quality. We employ methodology and tooling compatible with existing results in order to enable comparative evaluation. We carry out a long-term study targeting three complex, open-source applications, and provide the following contributions:
\begin{enumerate}[(i)]
    \item A clear description of our methodology, metric definitions and tooling used to extract metric values. Doing this ensures that our results can be used for comparative evaluation in future studies. We made all extracted metric values publicly available\footnote{\url{http://www.cs.ubbcluj.ro/~se/enase2019}}. 
    \item A quantitative evaluation of metric values is carried out and detailed for all target application versions.
    \item A longitudinal exploratory study that examines the evolution of metric values over the course of 18 years of target application development. 
    \item Identification of statistical correlations between metric pairs. We identify both strongly correlated metrics as well as metrics that appear independent. We account for the confounding effect of class size and examine the stability of the correlation strength across application versions.
    \item A comparative evaluation of metric values and statistical correlations between target applications. We identify trends in metric values and correlations that are application-specific, together with those that hold across the target applications.
    \item An evaluation of our obtained results in the context of existing research that uses the same methodology and software tools.
\end{enumerate}

One of our study's key contributions lies in the selection of target applications. Existing studies are built around one of the following two approaches. The first one is where a number of applications are selected, and for each of them several versions are studied \cite{11,49}. The second one considers a large number of target applications, that in many cases are automatically downloaded from open-source repositories \cite{24}, with a cross-sectional study including all of them \cite{24,47,50}. Our approach aims to complement existing research. We select a number of three open-source applications developed on the same platform, having comparable complexity and scope, and include all their released versions in our study. This results in a large number of application versions that ensures statistical significance. More so, our approach includes both initial application versions, which are sometimes very simple functionality-wise and bug-prone. We also include the latest application versions, that appear polished, have extensive features sets and a consistent user base. This enables us to study how metric values evolve together with the target applications, as well as to identify any existing trends that might be influenced by application development status.

Another important contribution regards careful selection of software metrics and extraction tools. As detailed in our initial evaluation \cite{60}, we selected the evaluated metrics in order to cover complexity, inheritance, coupling and cohesion \cite{5,18} as important characteristics of object-oriented software. In addition, the studied metrics can be found in existing literature studying software product quality \cite{47,49,50,60}. Selection of the right tools for metric value extraction is also important, as most metrics have more than one definition \cite{22,38}. As such, comparative evaluation can be carried out only with existing research that employs the same metrics, and that uses the same tooling to extract metric values. 

In our initial evaluation \cite{60}, we employed the VizzMaintenance tool\footnote{http://www.arisa.se/vizz\_analyzer.php}, as it provides formal definitions of the extracted metrics. In addition, using VizzMaintenance allows us to compare our results with those reported in \cite{24}, where authors use the same tool to carry out a cross-sectional study of 146 open-source applications. In our extensive literature survey, we identified \cite{24} as the only paper that clearly detailed the study methodology and tooling in order to allow a comparative evaluation to be carried out. Since our present paper employs the same methodology and tooling as our initial evaluation \cite{60}, the obtained results are directly comparable. In addition, in the present paper we explore the effect class size has on metric correlations across our target applications. We show that metric variability is greatest in early versions, before application architecture is well established. Furthermore, we find that most significant changes to metric values occur across a small number of application versions, which we examine in detail.

\section{Software Metrics}
\label{sec:metrics}
Evolution in the domain of software metrics was influenced by changes in the development of software, with increasingly specific metrics being proposed for the measurement of both software products as well as software processes. This is reflected in the appearance of software metric tools, both general and language dependent, stand-alone as well as integrated into IDEs in the form of plugins.

The oldest software metrics that remain widely used today include lines of code (LOC), number of functions or modules, and the number of comment lines. This was followed by proposed metrics to measure code complexity, such as cyclomatic complexity \cite{51} and Halstead volume \cite{52}. In turn, these were used to compute additional, more complex metrics such as the Maintainability Index \cite{20}. The object oriented paradigm introduced new entities and relations, and these were reflected by several newly proposed metrics. The reference set of object-oriented metrics was defined by Chidamber \& Kemerer (CK) \cite{6}, were implemented in most software metrics tools, and used in many subsequent studies. The lack of cohesion in methods (LCOM) metric deserves special mention, as it was refined from its original definition in \cite{6} by Li and Henry \cite{33}, and then by Hitz and Montazeri \cite{34}. While these changes were driven by a desire to better capture the essence of cohesion, LCOM values can only be compared when extracted using the same definition. Several tools are available to compute the CK metrics (and many more). Some of them are available as IDE plugins, such as Metrics2 \footnote{http://metrics.sourceforge.net} for Eclipse,  MetricsReloaded\footnote{https://plugins.jetbrains.com/plugin/93-metricsreloaded} for IntelliJ, NDepend \footnote{https://www.ndepend.com/} for .NET, or as standalone tools such as JHawk\footnote{http://www.virtualmachinery.com/jhawkprod.htm}, Sourcemeter \footnote{https://www.sourcemeter.com/}. Each of them employs its own implementation for metric computation, leading to different results for the same metric when extracted with different tools.

The metrics selected for our study were all computed using the VizzAnalyzer tool, that uses the definitions provided in \cite{62}. Other studies \cite{24,38} are based on the same tool, giving us the possibility to compare the obtained results. According to \cite{5}, object-oriented metrics measure one of the four internal characteristics essential to object orientation, namely coupling, inheritance, cohesion and structural complexity. We present the metrics used in our study, categorized according to the internal characteristics they aim to measure. We start with metrics dedicated to measuring \textbf{coupling}:
\begin{itemize}
\item \emph{Coupling Between Objects (\textbf{CBO}, $v_{CBO} \in [0,\infty) \cap \mathbb{Z}$)} \cite{14} - for class \textbf{c} is computed as the number of other classes that are coupled to it. Two classes are coupled when methods declared in one class use methods or instance variables defined by the other class. CBO indicates the required effort to test and maintain a class. 
\item \emph{Data Abstraction Coupling (\textbf{DAC}, $v_{DAC} \in [0,\infty) \cap \mathbb{Z}$)} \cite{33} - measures when a class is used in the implementation of methods of another class or when it is the domain of its instance variables. VizzAnalyzer does not include platform classes in this measurement.
\item \emph{Message Pass Coupling (\textbf{MPC}, $v_{MPC} \in [0,\infty) \cap \mathbb{Z}$)} \cite{14} - counts the number of methods from other classes that are called. It indicates the degree of dependency on the system's other classes.
\end{itemize}

The following metrics measure the \textbf{inheritance} characteristic:
\begin{itemize}
\item \emph{Depth of Inheritance Tree (\textbf{DIT}, $v_{DIT} \in [0,\infty) \cap \mathbb{Z}$)} \cite{14} - represents the length of the longest path from a given class to the root of the inheritance tree. DIT also accounts for multiple paths possible in the context of multiple-inheritance languages such as C++.
\item \emph{Number of Children (\textbf{NOC}, $v_{NOC} \in [0,\infty) \cap \mathbb{Z}$)} \cite{14,15} - counts the immediate subclasses found in the inheritance tree for a given class.
\end{itemize}

System \textbf{cohesion} is measured using the following metrics:
\begin{itemize}
\item \emph{Lack of Cohesion in Methods (\textbf{LCOM}, $v_{LCOM} \in [0,\infty) \cap \mathbb{Z}$)} \cite{14} - represents the difference between the number of methods pairs that don't have, respectively have, instance variables in common. This uses the original definition of the metric \cite{14}. 
\item \emph{Improvement to Lack of Cohesion in Methods (\textbf{ILCOM}, $v_{ILCOM} \in [1,\infty) \cap \mathbb{Z}$)} \cite{34} - this employs the improved definition provided by Hitz and Montazeri. In several papers and software tools this is referred to as LCOM5. 
\item \emph{Tight Class Cohesion (\textbf{TCC}, $v_{TCC} \in [0,1] \cap \mathbb{Q}$)} \cite{48} - defined as the ratio between the number of directly connected public methods in a class divided by the number of all possible connections between the public methods of that class.
\end{itemize}

We employ the following metrics that measure the \textbf{structural complexity} of classes:
\begin{itemize}
\item \emph{Locality of Data (\textbf{LD}, $v_{LD} \in [0,1] \cap \mathbb{Q}$)} \cite{34} - represents the ratio between the data that is local to a class and all the data used by the class. VizzAnalyzer includes non-public and inherited attributes.
\item \emph{Number of Attributes and Methods (\textbf{NAM}, $v_{NAM} \in [0,\infty) \cap \mathbb{Z}$)} \cite{14} - represents the total number of attributes and methods that are locally defined by the class. This includes static methods, but excludes constructors and inherited fields or methods.
\item \emph{Number of Methods (\textbf{NOM}, $v_{NOM} \in [0,\infty) \cap \mathbb{Z}$)} \cite{14} - represents the number of methods locally defined in the class. $NAM - NOM$ gives the number of locally defined attributes.
\item \emph{Response For a Class (\textbf{RFC}, $v_{RFC} \in [0,\infty) \cap \mathbb{Z}$)} \cite{14} - counts the number of methods that could be invoked as a response to a given message. RFC is the number of methods called by a given class.
\item \emph{Weighted Method Count (\textbf{WMC}, $v_{WMC} \in [0,\infty) \cap \mathbb{Z}$)} \cite{14} - defined as the sum of the complexities of all methods of a given class. The complexity of a method is its McCabe cyclomatic complexity \cite{51}.
\end{itemize}

Finally, we also examine metrics related with code \textbf{documentation}:
\begin{itemize}
    \item \textit{Length of Class Name (\textbf{LEN}, $v_{LEN} \in [1,\infty) \cap \mathbb{Z}$)} - the length of the class name counted in characters.
    \item \textit{Lack of Documentation (\textbf{LOD}, $v_{LOD} \in [0,1] \cap \mathbb{Q}$)} - the ratio of missing comments in a given class. Each class should have one comment per class, and an additional one for each defined method. This metric ignores the structure and the content of the comments.
\end{itemize}

Beside these metrics, we also measured the Lines of Code (LOC), since it is considered a universal software metric that can be used across most programming languages and which gives basic information about size of a project. The relation between object-oriented metrics and LOC is worthy of further investigation, especially as existing research showed that class size has a strong confounding influence on quality models based on metrics \cite{40}.

\section{State of the art}
The increasing attention given to software metrics is proven by the large number of studies in this domain. In most cases, existing research is geared towards one of the following three main directions: definition and analysis of proposed software metrics, software metric application in refactoring, and studying the relation between software metrics and software quality models.

\subsection{Metrics} 
New metrics are being defined in order to fine-tune the characteristics of software systems, and in order to better reflect the properties of source code and associated artefacts. Examples include approaches to improve estimation of the maintenance effort \cite{61}, in order to supersede existing measures such as the Maintainability Index \cite{20} which was shown to be outdated \cite{53,54,55}. Other studies propose new metrics to better capture system coupling or cohesion \cite{56,57}. 

Special interest has been also given to studying inter-metric dependency and correlation. A large scale study \cite{24} was carried out using 146 Java applications, with 16 metrics extracted using the VizzMaintenance tool. Barkman et al. applied different descriptive statistic techniques in order to detect metric dependencies. Landman et al. \cite{47} show that typical \textit{getters} and \textit{setters} can distort metric dependencies by artificially increasing dependency values. In \cite{40}, authors show that class size has a significant impact on metric correlation, using experimental data from a large scale telecommunication framework. These results illustrate that in order to validate strong conclusions derived from data analysis based on metric values, further research needs to be carried out. This is expected to be of special importance in the case of large-scale projects that were developed over a long period of time.

\subsection{Refactoring} 
One of the first applications of software metrics was to use the recorded values in order to detect design flaws that could be solved through refactoring.

The impact of four refactoring methods on several metrics is described in \cite{58}, based on the source code's abstract syntax tree representation. Another significant study \cite{59} refers to the impact 10 refactoring  methods have on  different  metrics, including the Maintainability  Index, cyclomatic  complexity,  DIT,  class coupling  and  LOC. Changes to maintainability and modifiability after refactoring are presented in \cite{59} through  an  empirical  evaluation. The experimental evaluations included in the aforementioned studies illustrate that, in the case of complex systems, refactoring plays an important role for easing maintenance and keeping system complexity under control. The decision of where and how to refactor can be taken based on extracted values of suitable software metrics.

\subsection{Software Quality Models}
\begin{figure}[h]
    \centering
    \includegraphics[width=\textwidth]{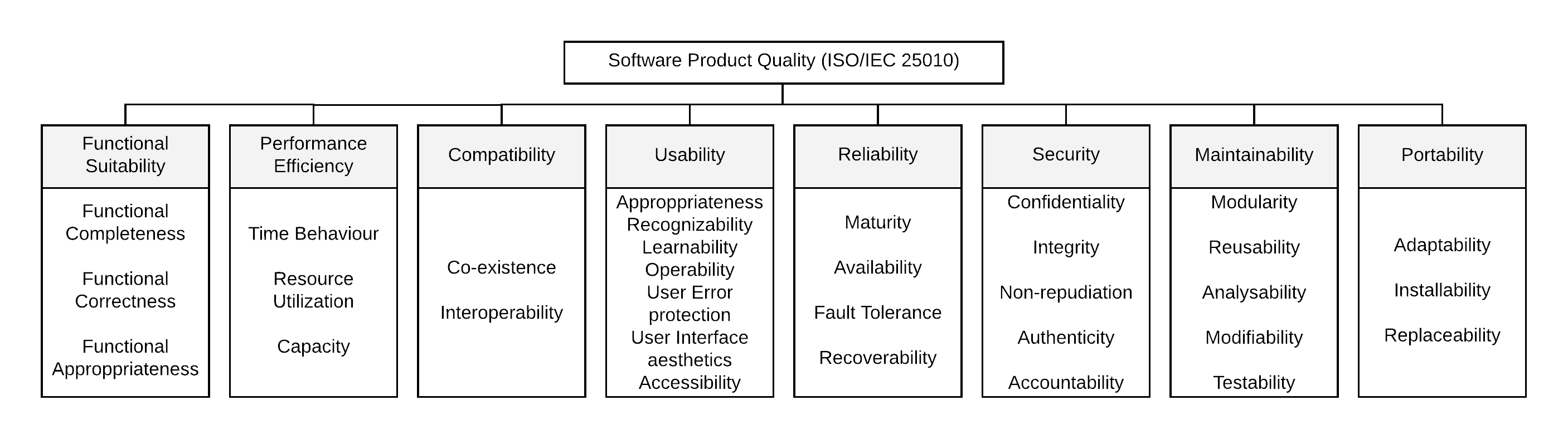}
    \caption{ISO/IEC 25010 subcharacteristics hierarchy}
    \label{fig:ISO25010}
\end{figure}

In recent years, several contributions attempted to connect software metrics with software quality factors. A software quality model is a hierarchical set of software quality factors or characteristics, that are further decomposed in subfactors or subcharacteristics. The first software quality model was introduced in 1976 by McCall, to which Boehm and Dromey later proposed important contributions. These initial contributions were later standardized by the ISO in the form of two families of standards: first, the ISO 9126, which expressed software quality as a function of six characteristics, that were comprised of 31 subcharacteristics. The 9126 standard was updated in the form of ISO 25010\footnote{
https://iso25000.com/index.php/en/iso-25000-standards/iso-25010}, which expands to the 8 characteristics shown in Figure \ref{fig:ISO25010}.

Some of the factors, like Maintainability, are known to be highly influenced by coupling and cohesion, such as evaluated by the CBO, TCC and LCOM metrics. However, in many other cases, dependencies remain to be proven. 

The ARISA Compendium \cite{18} offers an exhaustive study of the influence of over 20 metrics on the software quality characteristics of ISO 9126. The authors' approach is based on linking metrics with those source code entities that are involved in the metric's formal description. In \cite{33}, authors claim that metrics should be adapted for each programming paradigm. They introduce object oriented metrics for the maintenance effort and validate their approach on two commercial systems using 10 metrics. A complementary study was carried out in \cite{35}, where the CK metrics are assessed in regard to fault proneness, with experiments performed on eight C++ applications. The study concluded that LCOM, as defined in the CK suite is not evidential for fault detection, but that the other CK metrics are well suited for predicting faults. Also, the experimental data revealed an inverse relation between NOC and faults, a result confirmed also by the impact of reuse on fault proneness presented in \cite{45}. 

Another study \cite{44} regarding the relation between CK metrics and faults evaluated the efficient selection of testing techniques. Authors reported RFC and WMC as the most suited metrics for this task. A similar study was conducted in \cite{19} for the open-source Mozilla web and e-mail suite. It concluded that CBO and LOC are good predictors for faults, while DIT and NOC can lead to false results. An analysis \cite{36} of CK metrics on a NASA public data set revealed that LOC, WMC, CBO and RFC can be safely used for defect estimation. The conclusion of the study recommended further investigation on the relation between metric values and different dependent variables using statistical and AI techniques.

\section{Evaluation}
\subsection{Target Applications}
In order to carry out our evaluation, the first step was to select target applications. We first established several required criteria. First, we decided to target open-source applications developed in Java that were user interface driven and which did not have significant dependencies on external libraries or databases. We also searched for applications having long-term, consistent development history that were freely available. Our goals required a longitudinal study, an observational research method that consists in setting up and collecting metric data from each of the application versions. As detailed in \cite{24}, this can prove difficult in the case of open-source software, where development effort suffers interruptions, and where there are no guarantees that all software versions are complete and usable. As such, we selected three popular applications with long development histories, which had an established user base as well as public development repositories populated since project inception. We also ensured selected applications were free from complex dependencies. This allowed us to run them in order to check that functionalities worked as expected in all application versions. 

The selected applications are the FreeMind \footnote{http://freemind.sourceforge.net/wiki/index.php/Main\_Page} mind mapper, the jEdit \footnote{http://jedit.org} text editor and the TuxGuitar \footnote{http://www.tuxguitar.com.ar} tablature editor.  The entire development history of these applications can be found on SourceForge \footnote{https://sourceforge.net}.

\textbf{FreeMind} is a mind-mapping application that found many uses in productivity and content management. FreeMind was also employed in previous software research \cite{28}. It is also a popular application with a solid user base, having over 465k\footnote{Download data points taken on August 8\textsuperscript{th}, 2019.} downloads in 2019. FreeMind includes a plugin ecosystem with many plugins available. However, only the source code of the base application was included in our study.

\textbf{jEdit} is an open-source text editor, developed entirely using the Java programming language. It is also a popular system under test for other research endeavours in software testing \cite{28,29}. jEdit is one of the popular SourceForge applications, having over 59k downloads in 2019 and reaching over 8.9 millions downloads in its 19 years of existence. Similar to the case of FreeMind, plugin code was not included in our evaluation.

\textbf{TuxGuitar} is a free, open-source multitrack guitar tablature editor with an SWT-based user interface. It includes features like multiple format data import and export, tablature and score editing. TuxGuitar is also a popular application having over 131k downloads in 2019. In contrast with FreeMind and jEdit, where we disregarded the applications' plugin ecosystems, in the case of TuxGuitar functionalities related to data import and export itself were implemented in the form of a plugin, and were included in our evaluation.

Table \ref{tab:appevolution} provides information about the earliest and latest application versions included in our evaluation, indicating their change of complexity during the considered period.

\begin{table}[ht]
    \caption{First and last studied version of each target application (from \cite{60})}
    \label{tab:appevolution}
    \centering
    \begin{tabular}{|l|c|c|c|}
    \hline
         Application & Version & LOC & Classes \\
         \hline
         \multirow{2}{*}{jEdit} & 2.3pre2 & 33,768 & 322 \\
         \cline{2-4}
          & 5.5.0 & 151,672 & 952 \\
          \hline
          \multirow{2}{*}{FreeMind} & 0.0.3 & 3,722 & 53\\
          \cline{2-4}
          & 1.1.0Beta2 & 63,799 & 587\\
          \hline
          \multirow{2}{*}{TuxGuitar} & 0.1pre & 11,209 & 122\\
          \cline{2-4}
          & 1.5.2 & 108,495 & 1,618\\
          \hline
    \end{tabular}
\end{table}

\begin{figure}[!ht]
    \captionsetup[subfigure]{labelformat=empty,justification=centering}
    \centering
    \begin{subfigure}[b]{0.32\textwidth}
        \includegraphics[width=\textwidth]{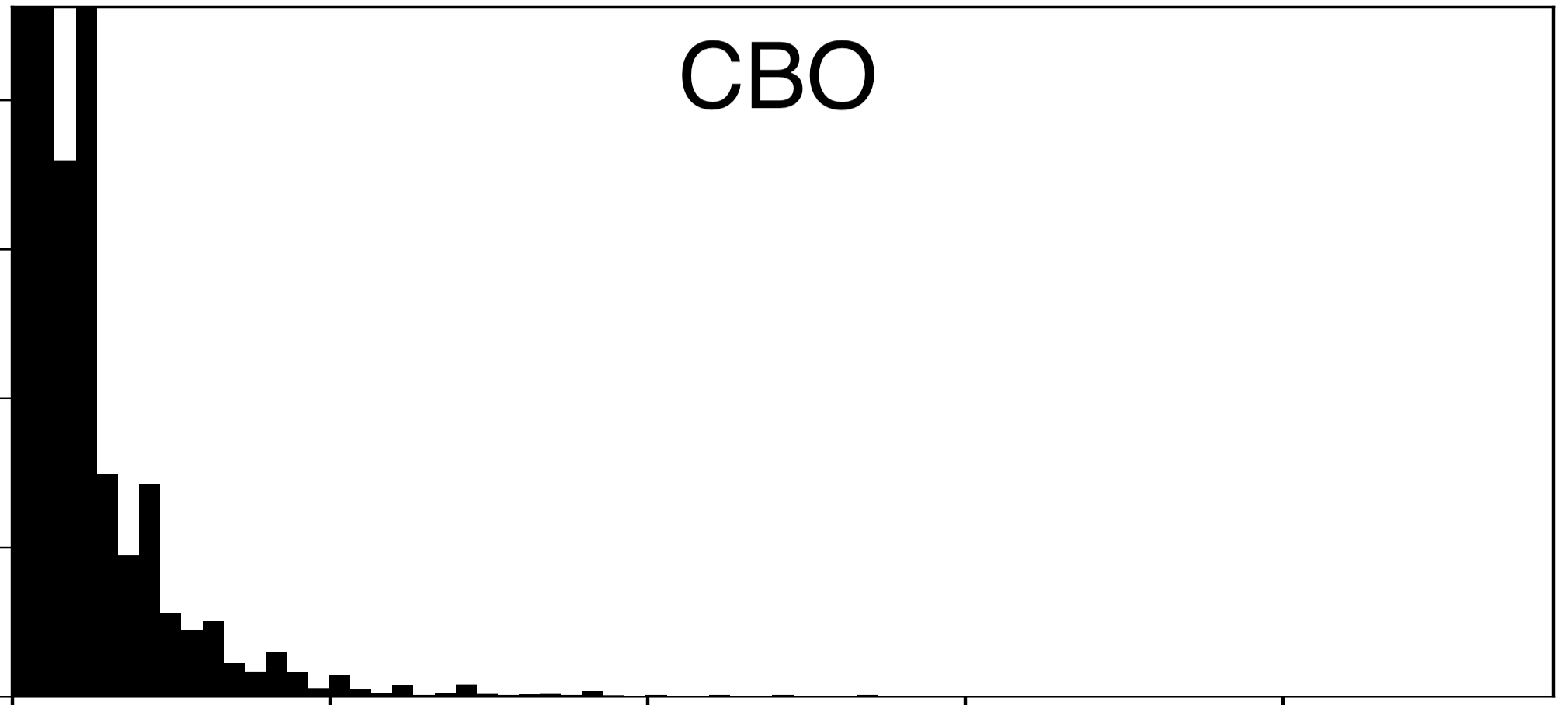}\subcaption{0, 5.61, 285, 3, 1\\0, 6.71, 184, 4, 1}
    \end{subfigure}
    \begin{subfigure}[b]{0.32\textwidth}
        \includegraphics[width=\textwidth]{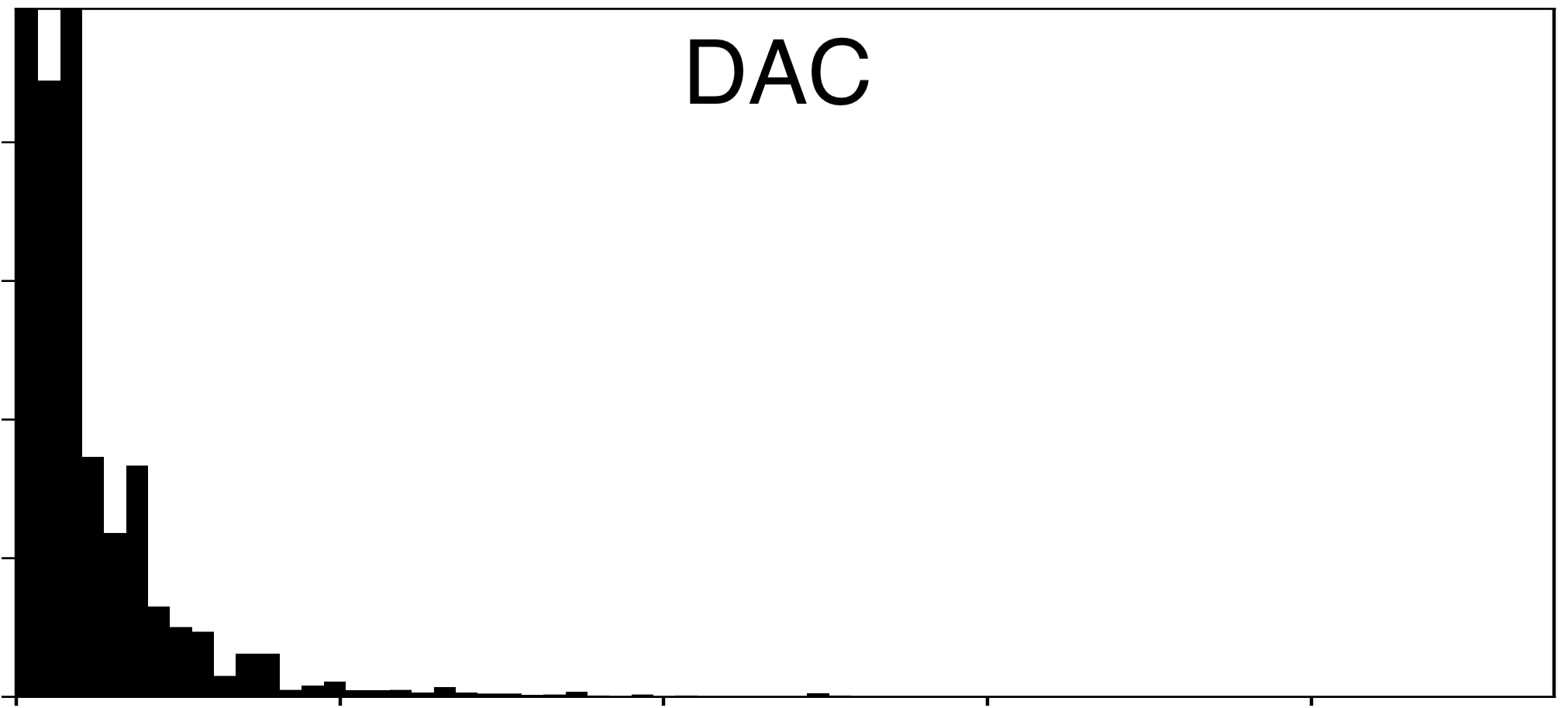}\subcaption{0, 4.62, 285, 3, 0\\0, 6.04, 175, 4, 1}
    \end{subfigure}
    \begin{subfigure}[b]{0.32\textwidth}
        \includegraphics[width=\textwidth]{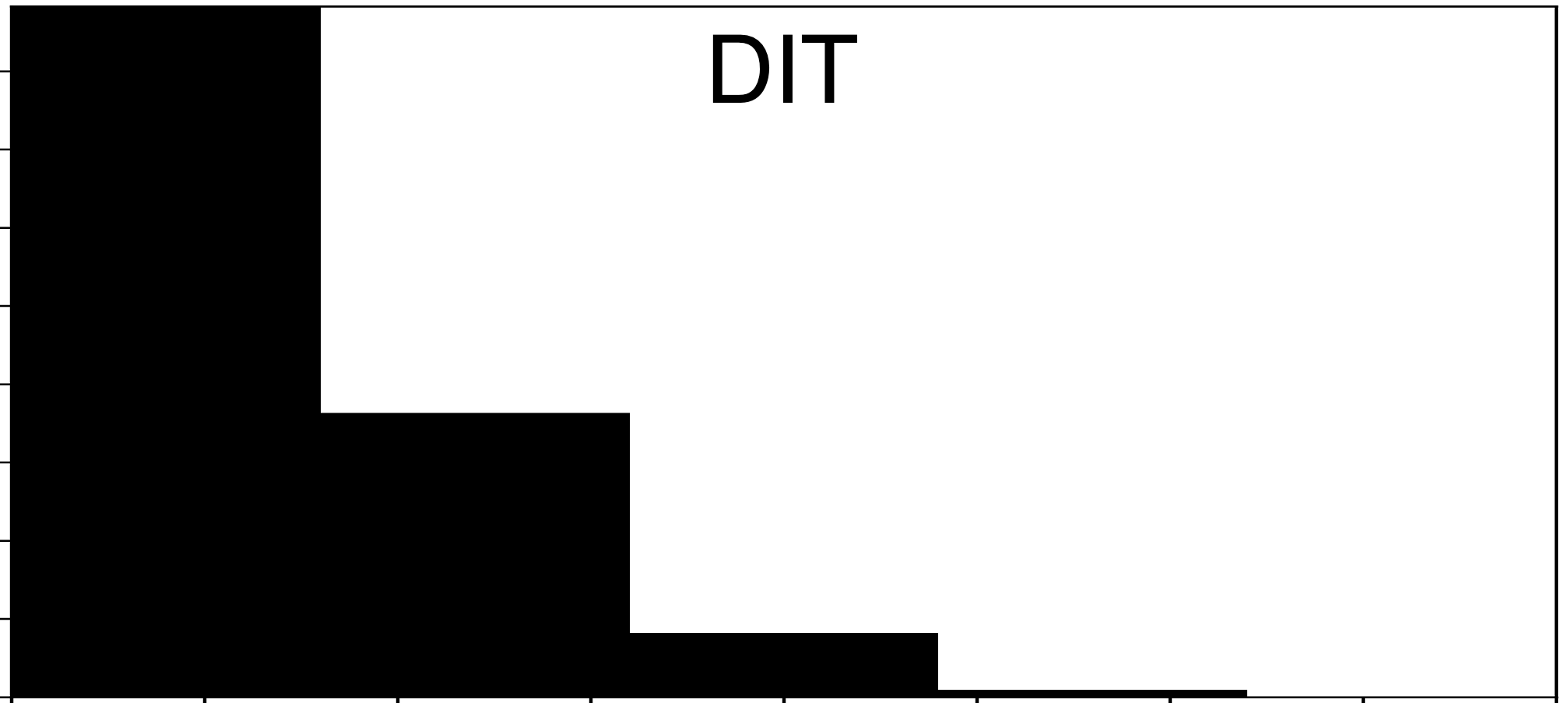}\subcaption{0, 0.65, 6, 0, 0\\0, 1.46, 9, 1, 0}
    \end{subfigure}
    \par\medskip

    \begin{subfigure}[b]{0.32\textwidth}
        \includegraphics[width=\textwidth]{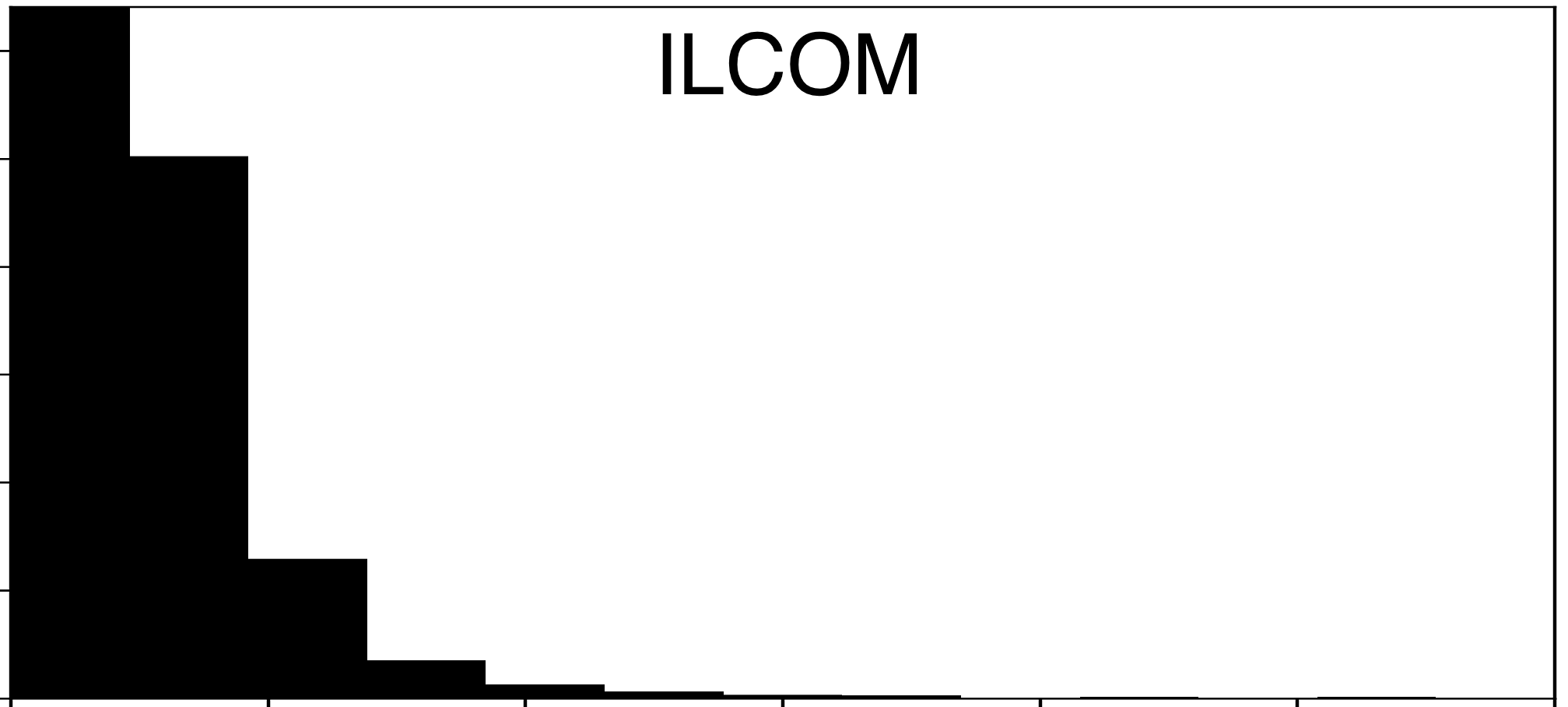}\subcaption{0, 0.97, 45, 1, 0\\0, 0.99, 298, 1, 0}
    \end{subfigure}
    \begin{subfigure}[b]{0.32\textwidth}
        \includegraphics[width=\textwidth]{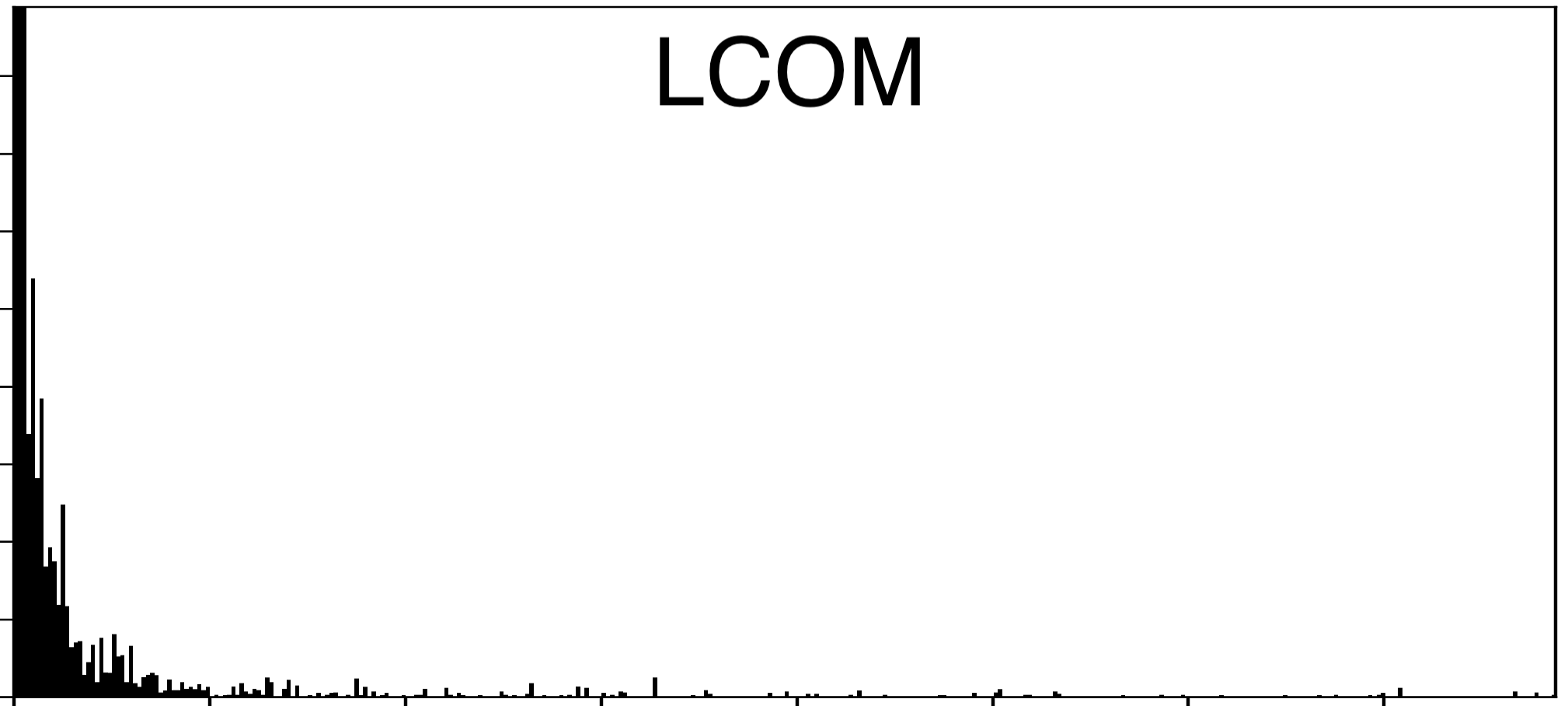}\subcaption{0, 147.41, 39813, 1, 0\\0, 210.11, 1415498, 2, 0}
    \end{subfigure}
    \begin{subfigure}[b]{0.32\textwidth}
        \includegraphics[width=\textwidth]{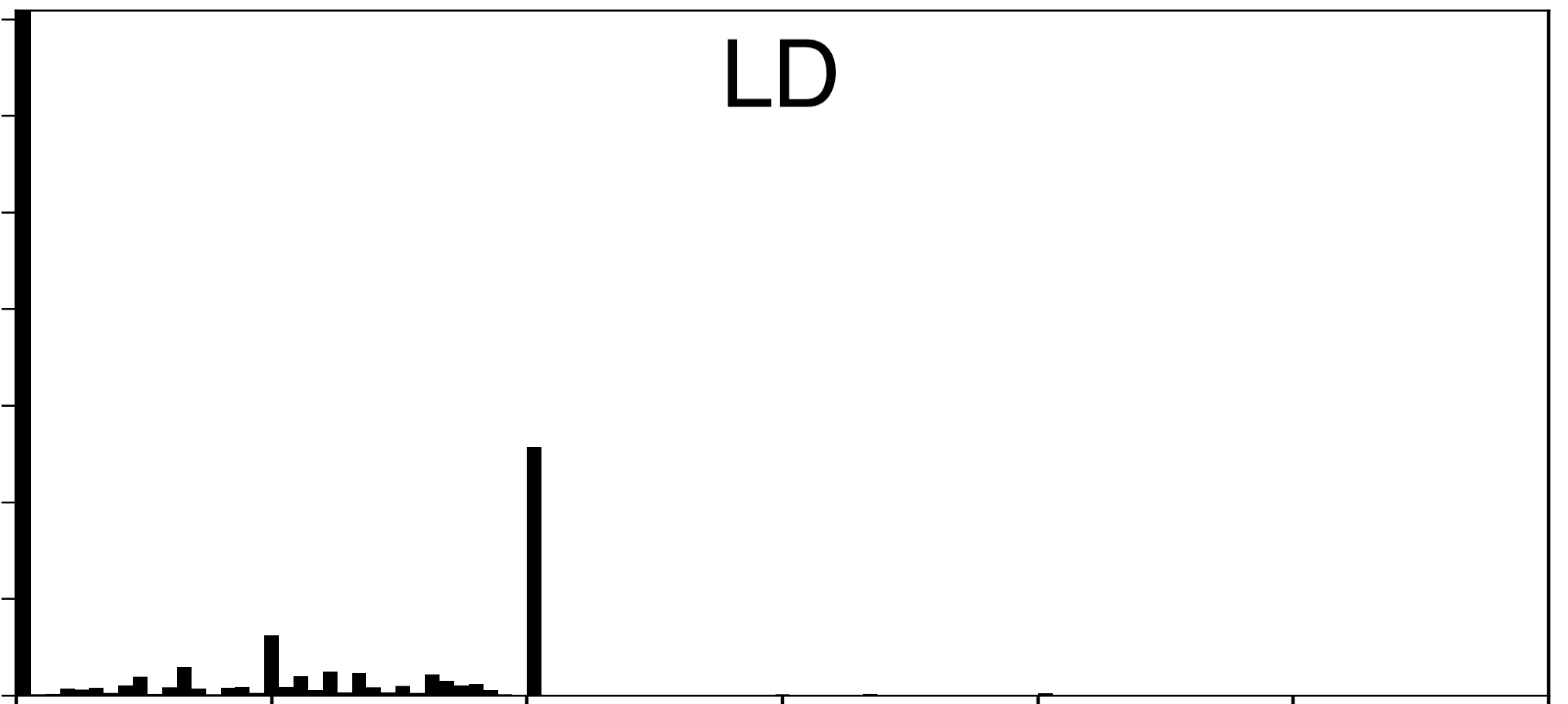}\subcaption{0, 0.4, 22, 0, 0\\0, 0.38, 36, 0, 0}
    \end{subfigure}
    \par\medskip

    \begin{subfigure}[b]{0.32\textwidth}
        \includegraphics[width=\textwidth]{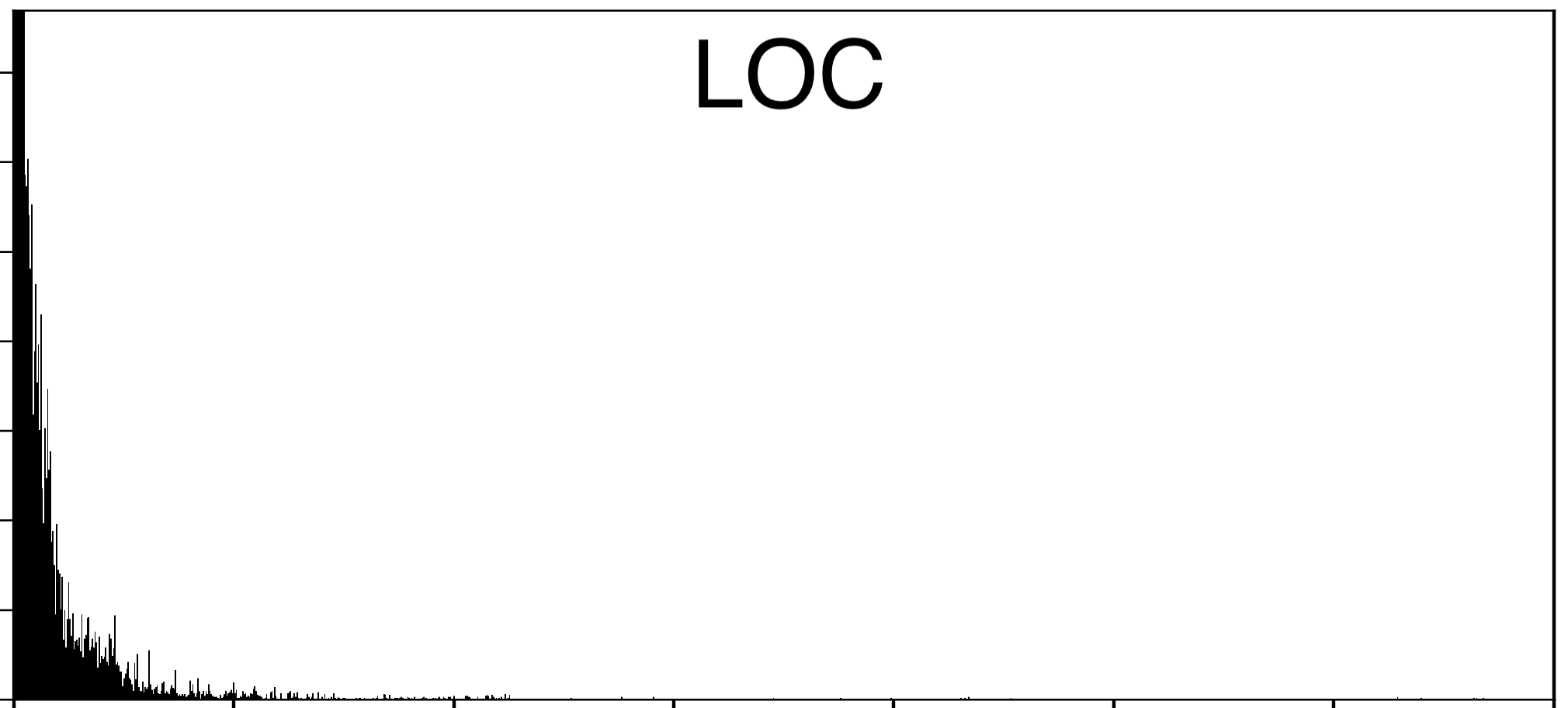}\subcaption{1, 124.37, 7001, 43, 13\\1, 166.9, 11045, 62, 13}
    \end{subfigure}
    \begin{subfigure}[b]{0.32\textwidth}
        \includegraphics[width=\textwidth]{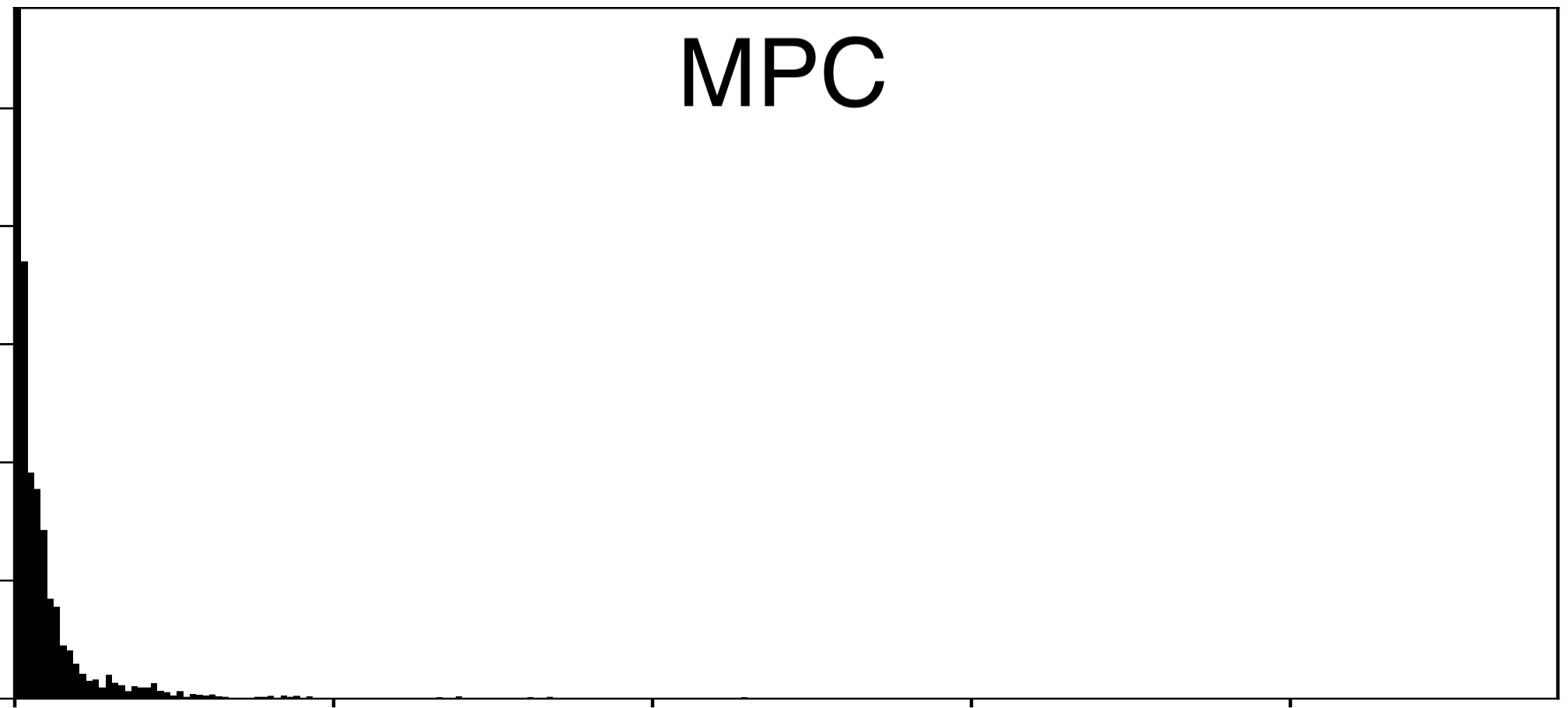}\subcaption{0, 12.12, 552, 4, 0\\0, 11.99, 1225, 3, 0}
    \end{subfigure}
    \begin{subfigure}[b]{0.32\textwidth}
        \includegraphics[width=\textwidth]{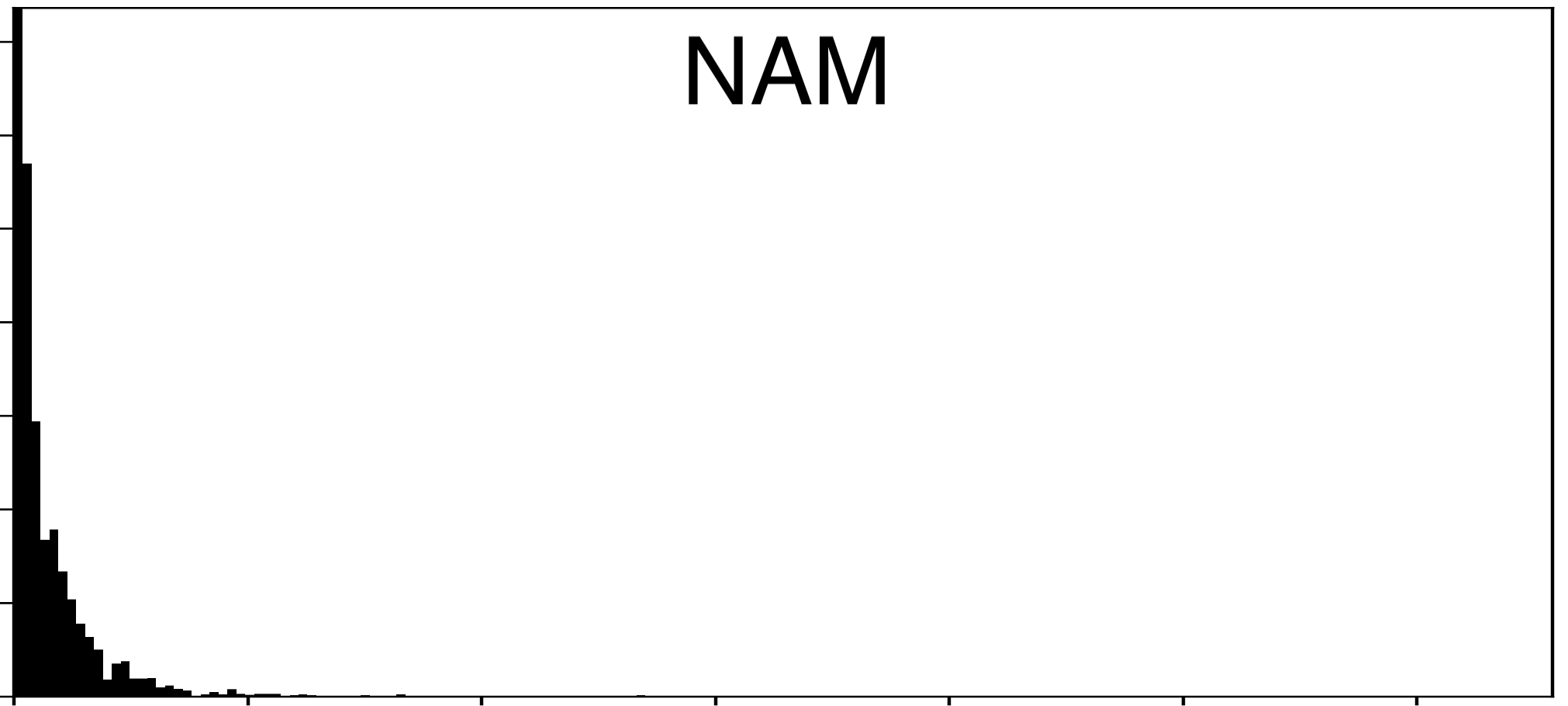}\subcaption{0, 9.42, 329, 4, 1\\0, 11.12, 2297, 5, 1}
    \end{subfigure}
    \par\medskip

    \begin{subfigure}[b]{0.32\textwidth}
        \includegraphics[width=\textwidth]{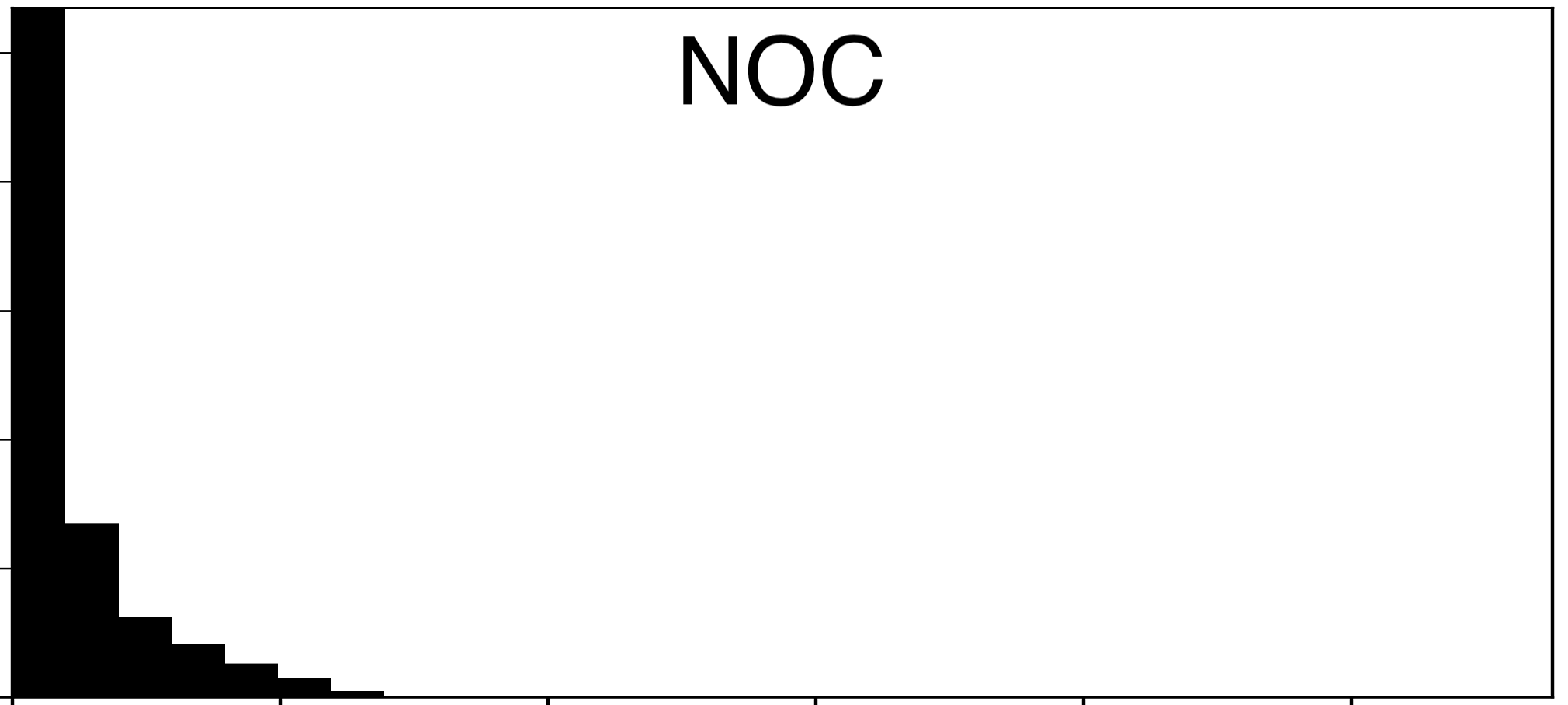}\subcaption{0, 0.54, 300, 0, 0\\0, 0.97, 2843, 0, 0}
    \end{subfigure}
    \begin{subfigure}[b]{0.32\textwidth}
        \includegraphics[width=\textwidth]{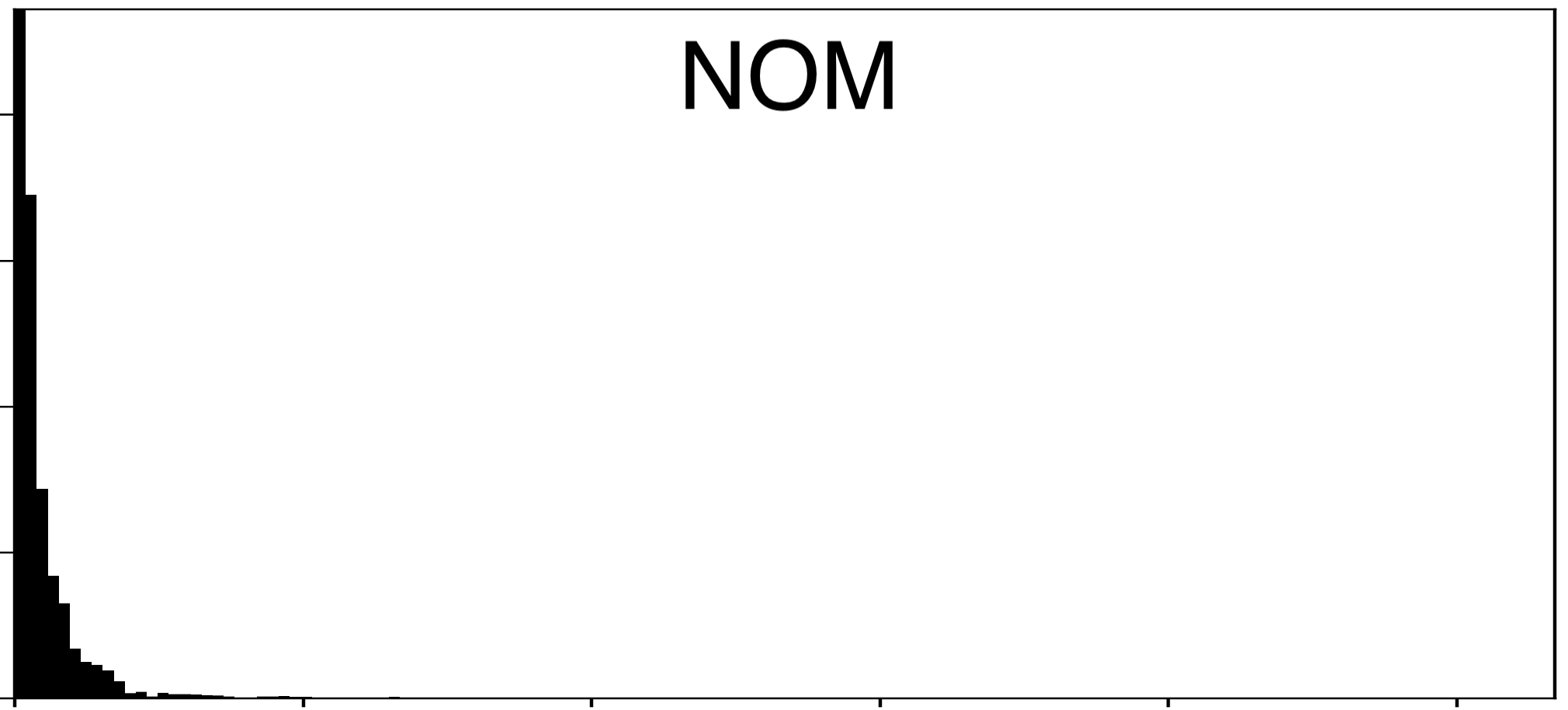}\subcaption{0, 6.11, 267, 3, 1\\0, 7.28, 1190, 3, 1}
    \end{subfigure}
    \begin{subfigure}[b]{0.32\textwidth}
        \includegraphics[width=\textwidth]{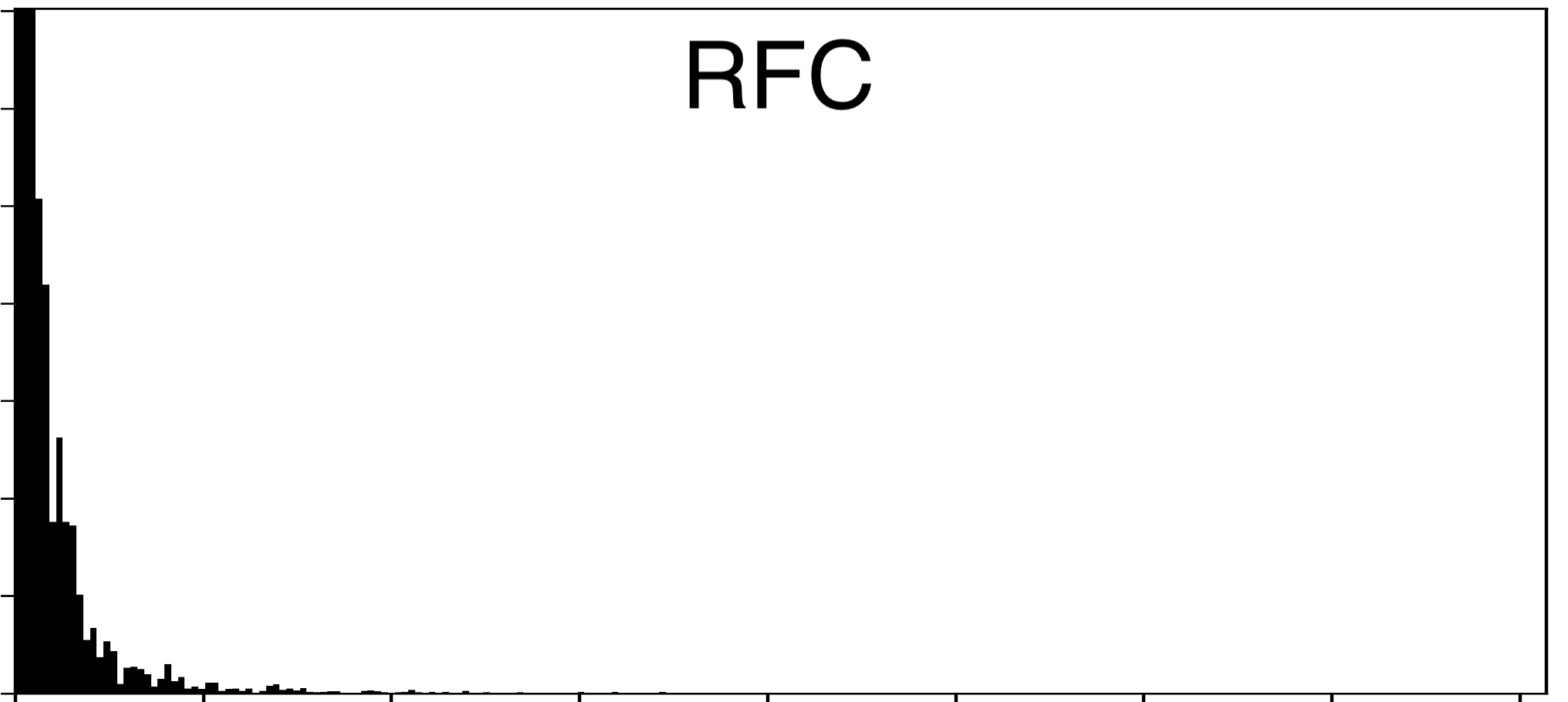}\subcaption{0, 13.46, 407, 6, 2\\0, 13.46, 1195, 6, 0}
    \end{subfigure}
    \par\medskip

    \begin{subfigure}[b]{0.32\textwidth}
        \includegraphics[width=\textwidth]{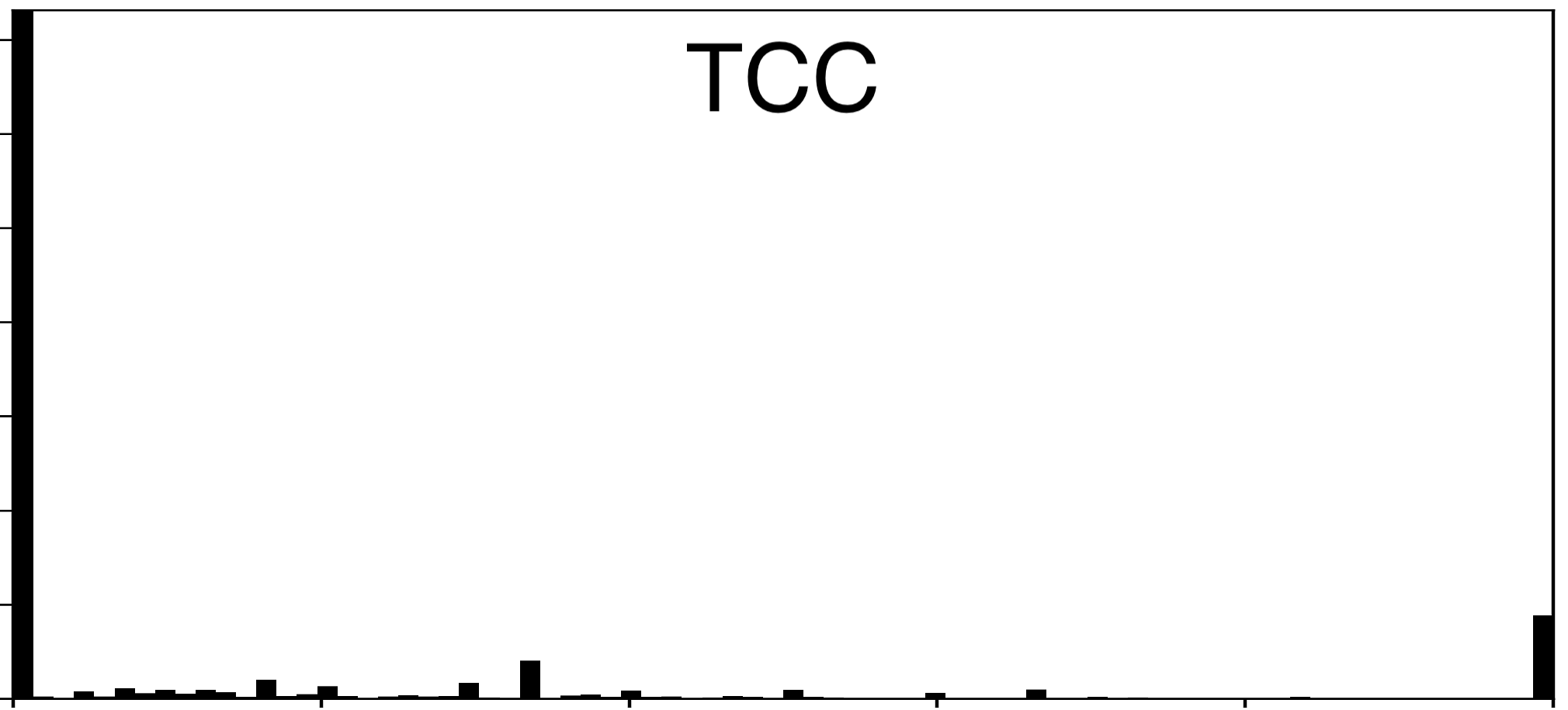}\subcaption{0, 0.14, 1, 0, 0\\0, 0.16, 1, 0, 0}
    \end{subfigure}
    \begin{subfigure}[b]{0.32\textwidth}
        \includegraphics[width=\textwidth]{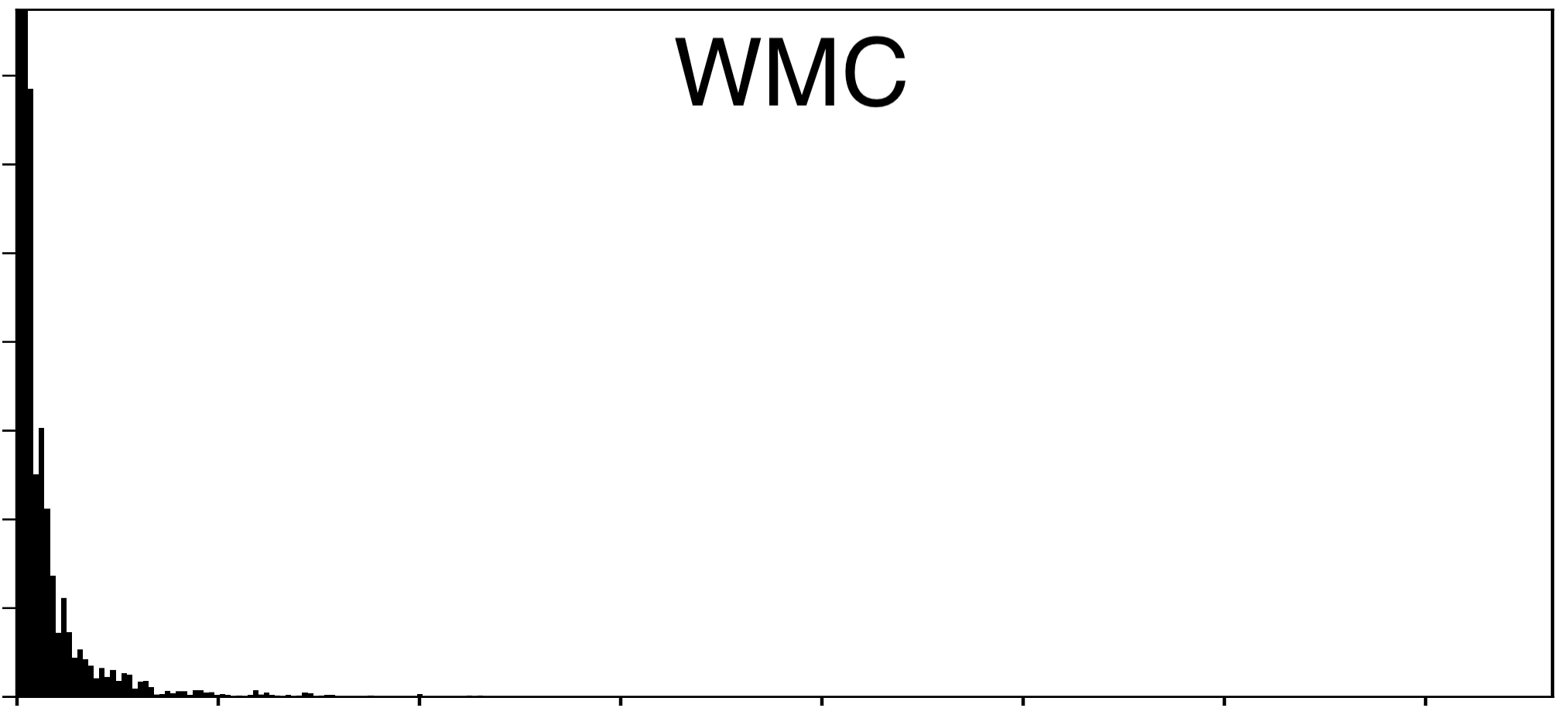}\subcaption{0, 12.85, 763, 5, 1\\ 0, 14.87, 2475, 5, 1}
    \end{subfigure}
    \par\medskip

    \caption{Code metric histograms. Data labels: minimum, mean, maximum, median, modus. Our results on top row, results from \cite{24} on bottom row for comparison (data from \cite{60}) }\label{fig:dscode}
\end{figure}

\begin{figure}[!ht]
    \captionsetup[subfigure]{labelformat=empty,justification=centering}
    \centering
    \begin{subfigure}[b]{0.32\textwidth}
        \includegraphics[width=\textwidth]{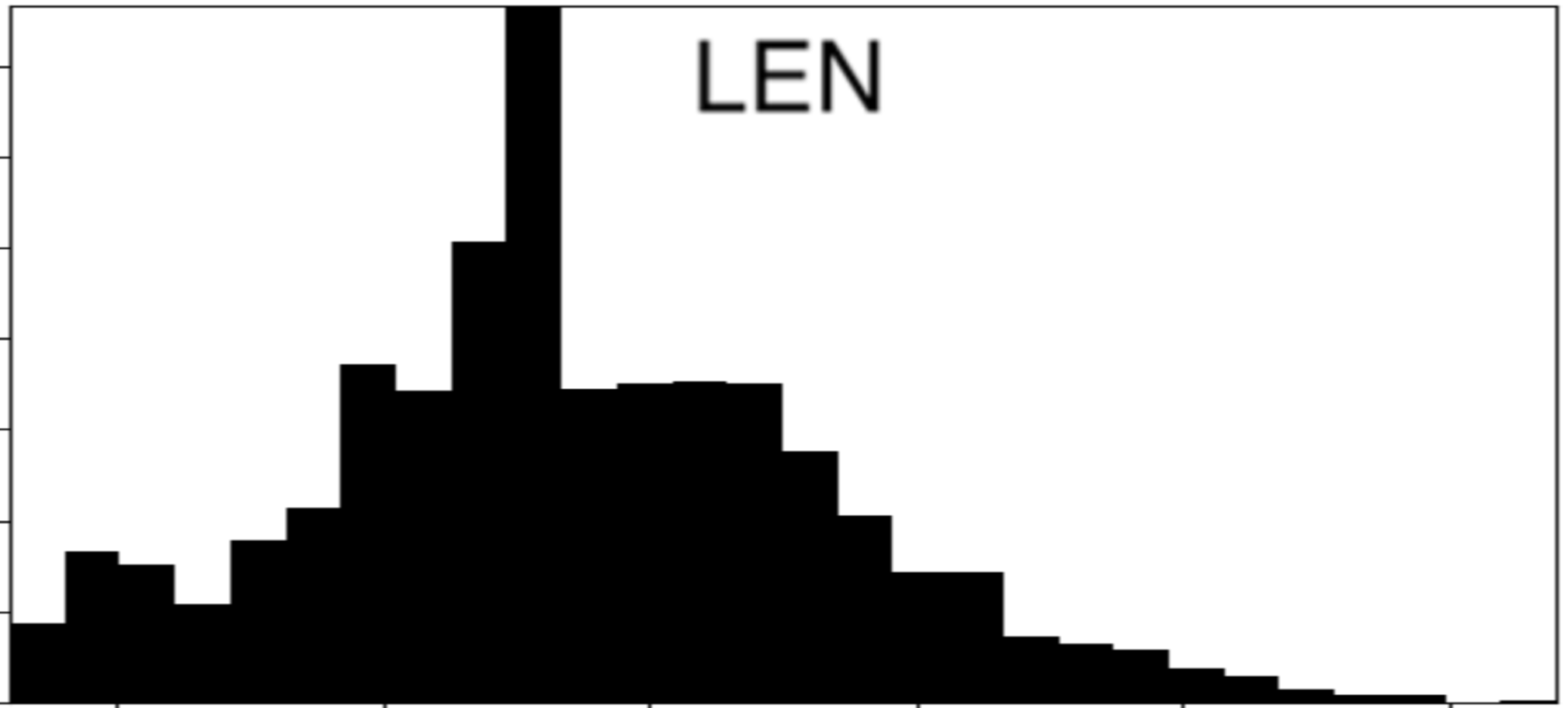}\subcaption{0, 5.61, 285, 3, 1\\0, 6.71, 184, 4, 1}
    \end{subfigure}
    \begin{subfigure}[b]{0.32\textwidth}
        \includegraphics[width=\textwidth]{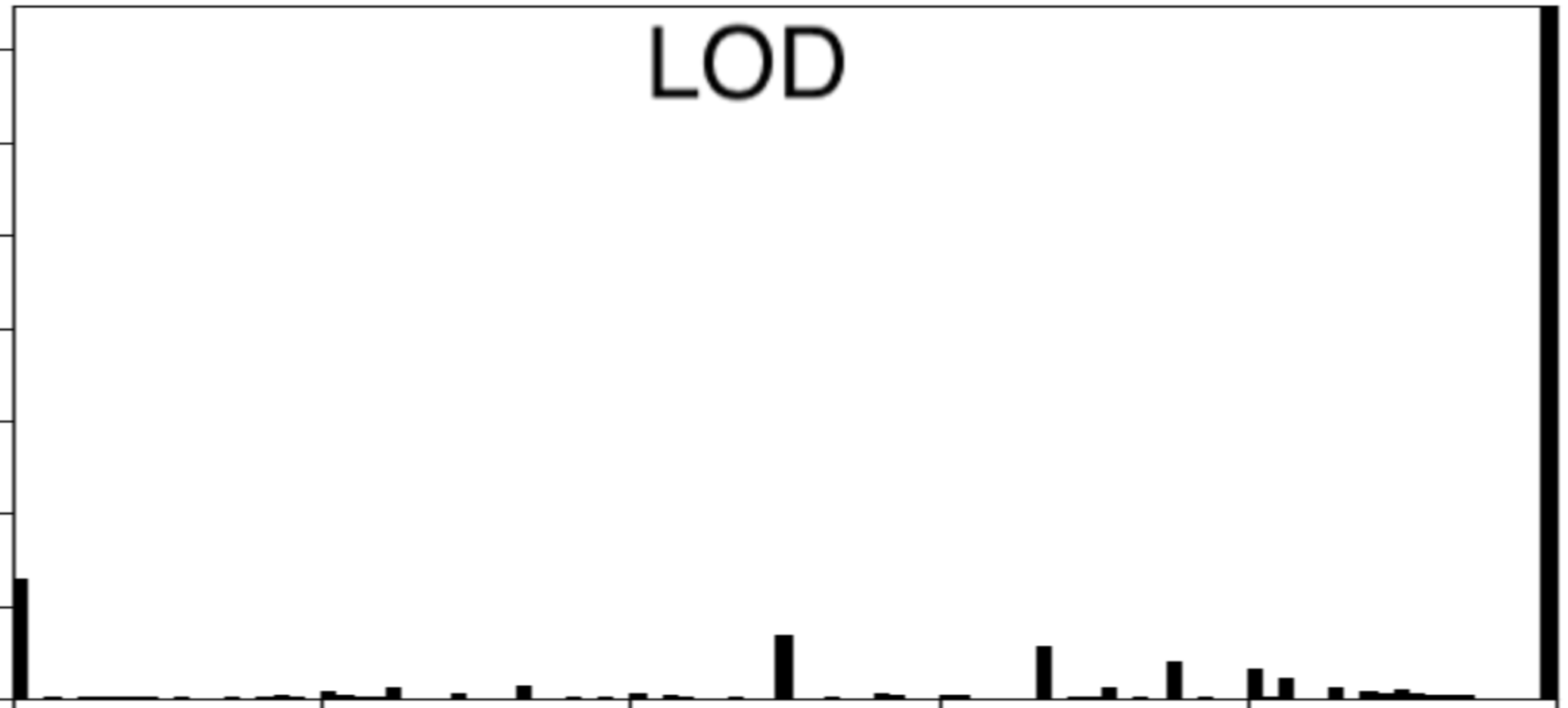}\subcaption{0, 4.62, 285, 3, 0\\0, 6.04, 175, 4, 1}
    \end{subfigure}
    \par\medskip

    \caption{Documentation metric histograms. Data labels: minimum, mean, maximum, median, modus. Our results on top row, results from \cite{24} on bottom row for comparison.}\label{fig:dsdoc}
\end{figure}

As a preparatory step, each studied version was imported into an IDE. We ensured that library source code was separated from actual application code in order to not affect our analysis. Since we employed Java 8, we encountered compilation errors with older versions of the applications that were developed using earlier versions of the Java platform. The issues were resolved taking into account not to alter the results of metric extraction. We assured that for each application, all mandatory source code was included, testing all available functionalities in detail. The raw metric data that was extracted is available on our website\footnote{\url{http://www.cs.ubbcluj.ro/~se/enase2019}}. Using this data, we developed a number of scripts in order to extract only the required metric values for our study for each application version as well as in aggregate form.

Data collection was helped by the fact that for each application, its complete development history was available on SourceForge. Furthermore, released versions were clearly marked, dated and had associated binaries and source code. In total, we included 38 versions of FreeMind, 43 for jEdit and 26 for TuxGuitar.

\subsection{Quantitative Statistics}
\label{sec:qs}
In this section we provide an initial overview of the extracted metric values, and compare them with the results presented in \cite{24}. For each of the target applications, we create its own data set, comprising metric values extracted from all studied versions of that application. This enables statistical comparison across applications in order to identify any existing trends. The data from all 107 application versions is coalesced into an aggregated data set. We compare the aggregated data against the results reported in \cite{24}, where authors carried out a cross-sectional study of 146 open-source Java applications. 

Given the large number of data points recorded for our study\footnote{107 application versions x 16 studied metrics x 5 data points = 8,560 data points.}, we detail those aspects that were found of most interest. We remind the interested reader that the entire metric data set is freely available on our website.

\begin{table*}
\caption{Mean and median metric values per application}\label{tab:dsmeans} \centering
\bgroup
\begin{tabular}{c c|c|c|c||c|c|c|c|}
    \cline{2-9} & \multicolumn{1}{|c|}{FreeMind} & jEdit & TuxGuitar & \cite{24} & FreeMind & jEdit & TuxGuitar & \cite{24} \\ \hline
    \multicolumn{1}{|c|}{CBO} & 5.36 & 4.67 & 7.32 & 6.71 & 3.00 & 3.00 & 5.00 & 4.00 \\
    \multicolumn{1}{|c|}{DAC} & 4.21 & 4.09 & 6.08 & 6.04 & 2.00 & 2.00 & 4.00 & 4.00 \\
    \multicolumn{1}{|c|}{DIT} & 0.79 & 0.42 & 0.87 & 1.46 & 0.00 & 0.00 & 1.00 & 1.00 \\
    \multicolumn{1}{|c|}{ILCOM} & 1.00 & 0.77 & 1.25 & 0.99 & 1.00 & 1.00 & 1.00 & 1.00 \\
    \multicolumn{1}{|c|}{LCOM} & 197.62 & 124.83 & 130.81 & 210.11 & 2.00 & 1.00 & 2.00 & 2.00 \\
    \multicolumn{1}{|c|}{LD} & 0.49 & 0.35 & 0.40 & 0.38 & 0.00 & 0.00 & 0.00 & 0.00 \\
    \multicolumn{1}{|c|}{LEN} & 16.87 & 13.67 & 16.88 & 15.04 & 16.00 & 13.00 & 16.00 & 14.00 \\
    \multicolumn{1}{|c|}{LOC} & 108.62 & 156.44 & 90.97 & 166.90 & 40.00 & 51.00 & 38.00 & 62.00 \\
    \multicolumn{1}{|c|}{LOD} & 0.80 & 0.76 & 0.92 & 0.47 & 1.00 & 1.00 & 1.00 & 0.50 \\
    \multicolumn{1}{|c|}{MPC} & 10.92 & 9.46 & 17.49 & 11.99 & 4.00 & 3.00 & 5.00 & 3.00 \\
    \multicolumn{1}{|c|}{NAM} & 9.75 & 8.41 & 10.67 & 11.12 & 4.00 & 3.00 & 5.00 & 5.00 \\
    \multicolumn{1}{|c|}{NOC} & 0.65 & 0.37 & 0.71 & 0.97 & 0.00 & 0.00 & 0.00 & 0.00 \\
    \multicolumn{1}{|c|}{NOM} & 6.88 & 5.16 & 6.80 & 7.28 & 3.00 & 2.00 & 3.00 & 3.00 \\
    \multicolumn{1}{|c|}{RFC} & 13.54 & 10.62 & 17.78 & 13.46 & 6.00 & 5.00 & 8.00 & 6.00 \\
    \multicolumn{1}{|c|}{TCC} & 0.14 & 0.15 & 0.16 & 0.16 & 0.00 & 0.00 & 0.00 & 0.00 \\
    \multicolumn{1}{|c|}{WMC} & 12.51 & 13.40 & 12.36 & 14.87 & 5.00 & 5.00 & 4.00 & 5.00 \\ \hline
    & \multicolumn{4}{c}{Mean values (\cite{60})} & \multicolumn{4}{c}{Median values} \\
\end{tabular}
\egroup
\end{table*}

Histograms for code and documentation metric values in our aggregated data set are shown in Figures \ref{fig:dscode} and \ref{fig:dsdoc}. They also provide a faithful representation of the value distributions from the three target application data sets. This also holds when comparing our data with that presented in \cite{24}. We find that histograms are similar even in the case of metrics having stand-out values, such as LD, LOD and TCC, where the value of 1 is frequent\footnote{In the case of TCC 1 is the maximal value}. LEN appears to be the only metric with normal distribution. 

Descriptive statistics for every metric in the aggregated data set, as well as corresponding ones from \cite{24} are shown below the histograms in Figures \ref{fig:dscode} and \ref{fig:dsdoc}. We notice that in every case, the smallest recorded values are the minimal ones, which is 0 for all metrics with the exception of LOC, where it is 1. Maximal values are outliers and show much more variance, both across studied application versions and across the data sets. As such, our study will focus mostly around median and mean metric values, and detail extreme values only where it makes sense.

Examination of the mean, median and modus values proves to be of much more interest. Our first observation is that median and modus values are close across all the five data sets, for each of the 16 studied metrics. This is detailed in Table \ref{tab:dsmeans}, where mean and median values for each application data set, as well as those recorded by Barkmann et al. \cite{24} are shown. When examining these values, one must also consider the range for each metric, as detailed in Section \ref{sec:metrics}. We observe that for CBO, NAM, NOM, TCC and WMC mean values are close across the data sets. Values for LEN and LOD show that while in most cases, the length of used identifiers is suitable, open-source applications appear to lack inline documentation. This is especially true in the case of our target applications, where more than 80\% of methods remain undocumented. The data also illustrates the existance of application-specific trends. We observe that jEdit classes tend to be larger, as illustrated by higher LOC than FreeMind and TuxGuitar, being very close to the mean LOC reported in \cite{24}. At the same time, jEdit shows a more flat inheritance hierarchy, illustrated by lower DIT and NOC values when compared to the other applications. As a matter of fact, our studied applications tend to have shallower inheritance trees than those from \cite{24}.

\subsection{Metric Dependencies}
Several metric value-based characterizations of software have been proposed in existing literature. However, many of them eschew a thorough study of the relations between numerical metric values. We believe that understanding existing correlations between metrics can further assist researchers in proposing and evaluating metric-based models. In this section we identify existing metric dependencies in the target applications and cross-check our data against \cite{24}. 

\begin{table*}[!t]
\caption{Metric dependencies in FreeMind (top row), jEdit (second row), TuxGuitar (third row) and as reported in \cite{24} (bottom row). LEN and LOD metrics omitted as no strong dependencies were found. Data from \cite{60}}\label{tab:correlations} \centering
\bgroup
\scriptsize
\begin{tabular}{|c|c|c|c|c|c|c|c|c|c|c|c|c|c|c|}
  \hline
  Metric & CBO & DAC & DIT & ILCOM & LCOM & LD & LOC & MPC & NAM & NOC & NOM & RFC & TCC & WMC\\  
  \hline
  \multirow{4}{*}{DAC} & \textbf{0.97} & & & & & & & & & & & & &\\
                       & \textbf{0.98} & 1.00 & & & & & & & & & & & &\\
                       & \textbf{0.96} & & & & & & & & & & & & &\\
                       & \textbf{0.98} & & & & & & & & & & & & &\\
  \hline
  \multirow{4}{*}{DIT} & 0.28 & 0.30 & & & & & & & & & & & &\\
                       & 0.18 & 0.20 & 1.00 & & & & & & & & & & &\\
                       & 0.18 & 0.10 & & & & & & & & & & & &\\
                       & 0.52 & 0.52 & & & & & & & & & & & &\\
  \hline
  \multirow{4}{*}{ILCOM} & 0.46 & 0.49 & 0.08 & & & & & & & & & & &\\
                         & 0.44 & 0.46 & -0.00 & 1.00 & & & & & & & & & &\\
                         & 0.07 & 0.11 & -0.29 & & & & & & & & & & &\\
                         & 0.53 & 0.41 & 0.39 & & & & & & & & & & &\\
  \hline
  \multirow{4}{*}{LCOM} & 0.53 & 0.56 & 0.05 & 0.55 & & & & & & & & & &\\
                        & 0.55 & 0.56 & -0.03 & 0.40 & 1.00 & & & & & & & & &\\
                        & 0.20 & 0.21 & -0.12 & 0.37 & & & & & & & & & &\\
                        & 0.53 & 0.55 & 0.40 & 0.47 & & & & & & & & & &\\
  \hline
  \multirow{4}{*}{LD}  & 0.20 & 0.22 & 0.07 & 0.40 & 0.11 & & & & & & & & &\\
                       & 0.18 & 0.21 & 0.15 & 0.56 & 0.07 & 1.00 & & & & & & & &\\
                       & 0.03 & 0.06 & -0.20 & 0.43 & 0.11 & & & & & & & & &\\
                       & 0.31 & 0.33 & 0.43 & 0.79 & 0.44 & & & & & & & & &\\
  \hline
  \multirow{4}{*}{LOC} & 0.58 & 0.61 & 0.09 & 0.56 & 0.77 & 0.25 & & & & & & & &\\
                       & 0.77 & 0.78 & -0.00 & 0.55 & \textbf{0.84} & 0.21 & 1.00 & & & & & & &\\
                       & 0.46 & 0.46 & -0.14 & 0.34 & 0.66 & 0.16 & & & & & & & &\\
                       & 0.58 & 0.60 & 0.14 & 0.47 & 0.58 & 0.32 & & & & & & & &\\
  \hline
  \multirow{4}{*}{MPC} & \textbf{0.83} & \textbf{0.81} & 0.22 & 0.46 & 0.60 & 0.17 & 0.66 & & & & & & &\\
                       & \textbf{0.83} & \textbf{0.82} & 0.06 & 0.44 & 0.75 & 0.15 & \textbf{0.87} & 1.00 & & & & & &\\
                       & 0.62 & 0.56 & 0.03 & 0.18 & 0.56 & 0.04 & \textbf{0.82} & & & & & & &\\
                       & \textbf{0.83} & \textbf{0.81} & 0.53 & 0.57 & 0.59 & 0.50 & 0.66 & & & & & & &\\
  \hline
  \multirow{4}{*}{NAM} & 0.69 & 0.72 & 0.11 & 0.72 & \textbf{0.86} & 0.32 & \textbf{0.85} & 0.71 & & & & & &\\
                       & 0.71 & 0.72 & -0.01 & 0.65 & \textbf{0.85} & 0.29 & \textbf{0.94} & \textbf{0.82} & 1.00 & & & & &\\
                       & 0.30 & 0.30 & -0.23 & 0.57 & 0.78 & 0.29 & 0.78 & 0.59 & & & & & &\\
                       & 0.51 & 0.53 & 0.16 & 0.63 & 0.68 & 0.46 & \textbf{0.83} & 0.62 & & & & & &\\
  \hline
  \multirow{4}{*}{NOC} & -0.01 & 0.02 & -0.03 & 0.10 & 0.14 & 0.01 & 0.06 & 0.02 & 0.13 & & & & &\\
                       & -0.04 & -0.03 & -0.05 & 0.02 & 0.01 & -0.01 & 0.01 & -0.02 & 0.01 & 1.00 & & & &\\
                       & -0.02 & -0.03 & -0.06 & 0.02 & 0.01 & 0.02 & -0.02 & -0.02 & 0.01 & & & & &\\
                       & 0.06 & 0.08 & 0.40 & 0.57 & 0.38 & 0.62 & -0.11 & 0.21 & 0.06 & & & & &\\
  \hline
  \multirow{4}{*}{NOM} & 0.56 & 0.60 & 0.10 & 0.65 & \textbf{0.91} & 0.23 & \textbf{0.82} & 0.63 & \textbf{0.95} & 0.16 & & & &\\
                       & 0.68 & 0.69 & -0.05 & 0.59 & \textbf{0.90} & 0.20 & \textbf{0.94} & \textbf{0.84} & \textbf{0.96} & 0.03 & 1.00 & & &\\
                       & 0.32 & 0.33 & -0.23 & 0.55 & \textbf{0.83} & 0.27 & \textbf{0.83} & 0.67 & \textbf{0.92} & 0.03 & & & &\\
                       & 0.56 & 0.58 & 0.23 & 0.59 & 0.79 & 0.48 & 0.79 & 0.65 & \textbf{0.91} & 0.14 & & & &\\
  \hline
  \multirow{4}{*}{RFC} & 0.74 & 0.74 & 0.18 & 0.62 & \textbf{0.84} & 0.23 & \textbf{0.80} & \textbf{0.88} & \textbf{0.91} & 0.11 & \textbf{0.90} & & &\\
                       & \textbf{0.83} & \textbf{0.82} & 0.02 & 0.53 & \textbf{0.82} & 0.18 & \textbf{0.92} & \textbf{0.96} & \textbf{0.91} & -0.01 & \textbf{0.93} & 1.00 & &\\
                       & 0.53 & 0.49 & -0.02 & 0.32 & 0.62 & 0.12 & \textbf{0.88} & \textbf{0.92} & 0.73 & -0.01 & \textbf{0.82} & & &\\
                       & 0.71 & 0.70 & 0.27 & 0.52 & 0.71 & 0.01 & \textbf{0.80} & \textbf{0.81} & \textbf{0.83} & 0.02 & \textbf{0.90} & & &\\
  \hline
  \multirow{4}{*}{TCC} & 0.02 & 0.02 & 0.02 & 0.11 & -0.04 & 0.22 & 0.03 & 0.04 & 0.05 & -0.02 & 0.02 & 0.04 & &\\
                       & 0.05 & 0.07 & 0.05 & 0.25 & -0.01 & 0.43 & 0.09 & 0.06 & 0.12 & -0.04 & 0.08 & 0.08 & 1.00 &\\
                       & 0.08 & 0.09 & -0.05 & 0.04 & -0.05 & 0.25 & 0.03 & -0.01 & 0.07 & -0.05 & 0.02 & 0.01 & &\\
                       & 0.33 & 0.35 & 0.54 & 0.78 & 0.46 & \textbf{0.80} & 0.26 & 0.51 & 0.41 & \textbf{0.84} & 0.45 & 0.36 & &\\
  \hline
  \multirow{4}{*}{WMC} & 0.53 & 0.55 & 0.08 & 0.61 & \textbf{0.86} & 0.23 & \textbf{0.89} & 0.69 & \textbf{0.90} & 0.12 & \textbf{0.93} & \textbf{0.90} & 0.04 &\\
                       & 0.70 & 0.70 & -0.04 & 0.53 & \textbf{0.88} & 0.16 & \textbf{0.95} & \textbf{0.87} & \textbf{0.93} & 0.01 & \textbf{0.96} & \textbf{0.93} & 0.08 & 1.00 \\
                       & 0.38 & 0.38 & -0.17 & 0.37 & 0.72 & 0.16 & \textbf{0.95} & \textbf{0.82} & 0.79 & -0.01 & \textbf{0.88} & \textbf{0.88} & 0.01 &\\
                       & 0.59 & 0.60 & 0.20 & 0.57 & 0.72 & 0.44 & \textbf{0.84} & 0.71 & \textbf{0.88} & 0.05 & \textbf{0.93} & \textbf{0.93} & 0.40 &\\
  \hline
\end{tabular}
\egroup
\end{table*}

As shown in Figures \ref{fig:dscode} and \ref{fig:dsdoc}, LEN is the only metric having a normal distribution. This, together with the difference in metric value ranges shown in Section \ref{sec:metrics}, determined us to employ Spearman's rank correlation to determine metric dependency. Correlation data per application, including results from \cite{24} are shown in Table \ref{tab:correlations}. We establish a threshold of 0.8 in absolute value for \textit{strong} correlations, which are highlighted and discussed below. In order to keep Table \ref{tab:correlations} readable, we did not include the LEN and LOD metrics, both of which appeared to be independent from other metrics as well as each other. The only exception is a weak correlation between DIT and LEN, which appeared in all studied applications, as well as \cite{24}. It is explained by the tendency of derived classes in inheritance hierarchies to have more detailed names than those of base classes or interfaces.

Metric correlations in our target applications follow the trends identified by Barkman et al. \cite{24}. We examine our results through the lens of the four characteristics of object-oriented software presented in Section \ref{sec:metrics}. 

We observe that strong and consistent correlations exist between coupling metrics CBO, DAC and MPC, as well as size-related metrics LOC, NAM and NOM. This was expected, as an increase in attributes or method count leads to increased class sizes when measured using metrics that predate object orientation. The same explanation covers the strong observed correlation between structural complexity RFC and WMC. 

The NOM metric is also correlated with LCOM and NAM. This confirms that an increased method count usually leads to a lack of cohesion. As the number of class methods is a part of the NAM metric, this correlation was also expected. Inheritance metrics DIT and NOC remain uncorrelated in all data sets, challenging the expectation that classes at the base of the inheritance tree have more children. 

An interesting result is that cohesion metrics LCOM, ILCOM and TCC do not show strong correlation in either of the studied data sets. LCOM shows a weak correlation with its improved variant in all data sets, showing that while they measure similar software aspects, there is enough differentiation between them. The result for TCC is more interesting, as the cross-sectional study in \cite{24} showed much stronger correlation than observed by us. We believe this is a result of target application selection, which highlights the necessity of backing up any metric-based model with exploratory evaluation.

\begin{table*}[t]
\caption{Metric dependencies in FreeMind (top row - below Q1, middle row - inter-quartile range, bottom row - above Q3) }
\label{tab:corr_freemind} \centering
\bgroup
\scriptsize
\begin{tabular}{|c|c|c|c|c|c|c|c|c|c|c|c|c|c|}
\hline
Metric & CBO & DAC & DIT & ILCOM & LCOM & LD & MPC & NAM & NOC & NOM & RFC & TCC & WMC \\
  \hline
  \multirow{3}{*}{DAC} & 0.68 &  &  &  &  &  &  &  &  &  &  & &\\
                       & \textbf{0.83} & 1.00 &  &  &  &  &  &  &  &  &  & &\\
                       & \textbf{0.99} &  &  &  &  &  &  &  &  &  &  & &\\
  \hline
  \multirow{3}{*}{DIT} & 0.34 & 0.59 &  &  &  &  &  &  &  &  &  & &\\
                      & 0.42 & 0.55 & 1.00 &  &  &  &  &  &  &  &  & &\\
                      & 0.26 & 0.25 &  &  &  &  &  &  &  &  &  & &\\
  \hline
  \multirow{3}{*}{ILCOM} & 0.04 & 0.00 & -0.10 &  &  &  &  &  &  &  &  & &\\
                         & 0.04 & 0.17 & -0.03 & 1.00 &  &  &  &  &  &  &  & &\\
                         & 0.37 & 0.40 & 0.00 &  &  &  &  &  &  &  &  & &\\
  \hline
  \multirow{3}{*}{LCOM} & -0.10 & -0.05 & -0.12 & -0.20 &  &  &  &  &  &  &  & &\\
                        & 0.08 & 0.20 & 0.07 & -0.01 & 1.00 &  &  &  &  &  &  & &\\
                        & 0.49 & 0.51 & -0.01 & 0.55 &  &  &  &  &  &  &  & &\\
  \hline
  \multirow{3}{*}{LD} & 0.09 & 0.04 & -0.07 & \textbf{0.88} & -0.16 &  &  &  &  &  &  & &\\
                      & 0.16 & 0.25 & 0.08 & 0.66 & -0.07 & 1.00 &  &  &  &  &  & &\\
                      & 0.04 & 0.06 & -0.08 & 0.18 & 0.02 &  &  &  &  &  &  & &\\
  \hline
  \multirow{3}{*}{MPC} & 0.67 & 0.21 & -0.07 & 0.30 & -0.13 & 0.36 &  &  &  &  &  & &\\
                       & 0.78 & 0.59 & 0.32 & -0.03 & -0.05 & 0.14 & 1.00 &  &  &  &  & &\\
                       & \textbf{0.81} & 0.79 & 0.20 & 0.36 & 0.56 & -0.01 &  &  &  &  &  & &\\
  \hline
  \multirow{3}{*}{NAM} & -0.12 & -0.12 & -0.29 & 0.37 & 0.69 & 0.27 & -0.03 &  &  &  &  & &\\
                       & 0.13 & 0.32 & 0.10 & 0.53 & 0.64 & 0.41 & 0.03 & 1.00 &  &  &  & &\\
                       & 0.65 & 0.68 & -0.01 & 0.66 & \textbf{0.89} & 0.18 & 0.65 &  &  &  &  & &\\
  \hline
  \multirow{3}{*}{NOC} & -0.19 & -0.04 & -0.11 & -0.11 & 0.33 & -0.10 & -0.26 & 0.26 &  &  &  & &\\
                      & -0.10 & -0.04 & -0.08 & -0.03 & 0.22 & -0.03 & -0.17 & 0.19 & 1.00 &  &  & &\\
                      & 0.02 & 0.02 & 0.08 & 0.22 & 0.21 & 0.04 & 0.05 & 0.18 &  &  &  & &\\
  \hline
  \multirow{3}{*}{NOM} & -0.08 & -0.09 & -0.25 & 0.07 & \textbf{0.84} & 0.00 & -0.07 & \textbf{0.88} & 0.33 &  &  & &\\
                       & 0.16 & 0.32 & 0.14 & 0.27 & \textbf{0.80} & 0.21 & 0.04 & \textbf{0.90} & 0.24 & 1.00 &  & &\\
                       & 0.48 & 0.50 & -0.04 & 0.59 & \textbf{0.95} & 0.07 & 0.55 & \textbf{0.94} & 0.23 &  &  & &\\
  \hline
  \multirow{3}{*}{RFC} & 0.50 & 0.07 & -0.24 & 0.23 & 0.44 & 0.24 & 0.71 & 0.53 & 0.02 & 0.57 &  & &\\
                       & 0.66 & 0.61 & 0.29 & 0.15 & 0.45 & 0.25 & 0.74 & 0.58 & 0.06 & 0.63 & 1.00 & &\\
                       & 0.68 & 0.68 & 0.09 & 0.54 & \textbf{0.86} & 0.04 & \textbf{0.85} & \textbf{0.88} & 0.18 & \textbf{0.88} &  & &\\
  \hline
  \multirow{3}{*}{TCC} & -0.13 & -0.07 & 0.00 & 0.48 & -0.11 & 0.27 & -0.12 & 0.32 & -0.03 & 0.18 & 0.00 & &\\
                      & -0.01 & 0.07 & 0.01 & 0.31 & -0.14 & 0.35 & 0.05 & 0.27 & 0.00 & 0.14 & 0.16 & 1.00 &\\
                      & -0.10 & -0.12 & -0.10 & -0.15 & -0.18 & 0.07 & -0.09 & -0.17 & -0.03 & -0.20 & -0.16 & &\\
  \hline
  \multirow{3}{*}{WMC} & 0.09 & -0.01 & -0.23 & 0.04 & 0.76 & 0.01 & 0.08 & 0.78 & 0.27 & \textbf{0.90} & 0.65 & 0.12 &\\
                       & 0.23 & 0.33 & 0.09 & 0.22 & 0.59 & 0.24 & 0.23 & 0.77 & 0.13 & \textbf{0.83} & 0.70 & 0.18 & 1.00\\
                       & 0.42 & 0.43 & -0.06 & 0.53 & \textbf{0.88} & 0.06 & 0.60 & \textbf{0.86} & 0.18 & \textbf{0.91} & \textbf{0.87} & -0.13 &\\
\hline
\end{tabular}
\egroup
\end{table*}

\subsection{The confounding effect of class size}
The confounding effect class size has on metric value-based measurements was reported by El Emam et al. \cite{40}. Due to its significance, class size must be accounted for when studying metric dependencies. Authors of \cite{40} showed that in many cases, metric dependencies could be explained by larger classes having higher metric values, which confounds data interpretation. As shown in Table \ref{tab:correlations}, the LOC metric appears correlated with most of the metrics. The exceptions are DIT, LEN, LOD, NOC and TCC, which do not exhibit correlation with LOC, or other metrics.

To determine the effect class size has on metric dependencies, we partitioned all analyzed classes into quartiles using the LOC metric. We calculated the metric dependencies for each of our three data sets below the first quartile (below Q1), between the quartiles, and above the third quartile (above Q3). The detailed result is illustrated per application in Tables \ref{tab:corr_freemind}, \ref{tab:corr_jedit} and \ref{tab:corr_tuxguitar}. The LOC metric itself was omitted, as we had already used it to partition the data. 

Immediately we observe that most of the strong metric dependencies occur in classes above the third quartile, which confirms El Emam et al.'s observation of the important role played by class size in metric dependencies. LCOM, NAM and RFC appear sensitive to class size across all target applications, showing strong dependencies for classes above Q3. An inverse relation is observed between DIT on one hand, and CBO and DAC on the other. In this case, we notice dependency strength decrease for larger class sizes. This is to be expected, as most metrics capture state and behaviour introduced by the class itself, disregarding inherited attributes. As such, many classes deep in inheritance hierarchies appear deceptively simple, as much of their complexity is hidden in base classes.

Even with class size accounted for, we still observe highly dependent metric pairs. Coupling metrics CBO and DAC, as well as complexity metrics NOM and WMC illustrate this best. In the same way, metric pairs that we observed to be independent in the previous section remain so even when partitioned according to class size. DIT, NOC and TCC showed no strong dependency in any of the data partitions.

\begin{table*}[t]
\caption{Metric dependencies in jEdit (top row - below Q1, middle row - inter-quartile range, bottom row - above Q3) }
\label{tab:corr_jedit} \centering
\bgroup
\scriptsize
\begin{tabular}{|c|c|c|c|c|c|c|c|c|c|c|c|c|c|}
\hline
Metric & CBO & DAC & DIT & ILCOM & LCOM & LD & MPC & NAM & NOC & NOM & RFC & TCC & WMC \\
  \hline
  \multirow{3}{*}{DAC} & \textbf{0.82} & &  &  &  &  &  &  &  &  &  & &\\
                       & \textbf{0.94} & 1.00 &  &  &  &  &  &  &  &  &  & &\\
                       & \textbf{0.99} & &  &  &  &  &  &  &  &  &  & &\\
  \hline
  \multirow{3}{*}{DIT} & 0.34 & 0.50 & &  &  &  &  &  &  &  &  & &\\
                       & 0.33 & 0.39 & 1.00 &  &  &  &  &  &  &  &  & &\\
                       & 0.09 & 0.09 & &  &  &  &  &  &  &  &  & &\\
  \hline
  \multirow{3}{*}{ILCOM} & -0.06 & 0.01 & 0.05 &  &  &  &  &  &  &  &  & &\\
                         & -0.08 & -0.02 & -0.02 & 1.00 &  &  &  &  &  &  &  & &\\
                         & 0.36 & 0.36 & -0.13 &  &  &  &  &  &  &  &  & &\\
  \hline
  \multirow{3}{*}{LCOM} & 0.07 & 0.08 & -0.06 & -0.12 &  &  &  &  &  &  &  & &\\
                        & -0.04 & -0.02 & -0.02 & 0.11 & 1.00 &  &  &  &  &  &  & &\\
                        & 0.61 & 0.62 & -0.09 & 0.47 &  &  &  &  &  &  &  & &\\
  \hline
  \multirow{3}{*}{LD} & -0.10 & -0.02 & 0.04 & 0.78 & -0.10 &  &  &  &  &  &  & &\\
                      & -0.11 & -0.04 & 0.11 & 0.66 & -0.03 & 1.00 &  &  &  &  &  & &\\
                      & -0.02 & -0.01 & 0.16 & 0.23 & 0.00 &  &  &  &  &  &  & &\\
  \hline
  \multirow{3}{*}{MPC} & 0.68 & 0.45 & 0.11 & -0.09 & 0.01 & -0.08 &  &  &  &  &  & &\\
                       & 0.74 & 0.66 & 0.28 & -0.03 & -0.04 & 0.01 & 1.00 &  &  &  &  & &\\
                       & \textbf{0.84} & \textbf{0.83} & 0.01 & 0.40 & 0.78 & -0.03 &  &  &  &  &  & &\\
  \hline
  \multirow{3}{*}{NAM} & -0.15 & -0.05 & -0.21 & 0.40 & 0.39 & 0.38 & -0.22 &  &  &  &  & &\\
                       & -0.10 & -0.02 & 0.00 & 0.62 & 0.34 & 0.49 & -0.01 & 1.00 &  &  &  & &\\
                       & 0.70 & 0.70 & -0.09 & 0.59 & \textbf{0.91} & 0.07 & \textbf{0.82} &  &  &  &  & &\\
  \hline
  \multirow{3}{*}{NOC} & -0.17 & -0.08 & -0.11 & -0.08 & 0.16 & -0.06 & -0.21 & 0.04 &  &  &  & &\\
                       & -0.11 & -0.07 & -0.07 & -0.04 & 0.11 & -0.01 & -0.14 & 0.00 & 1.00 &  &  & &\\
                       & -0.07 & -0.06 & -0.01 & 0.07 & 0.02 & 0.01 & -0.04 & 0.00 &  &  &  & &\\
  \hline
  \multirow{3}{*}{NOM} & 0.15 & 0.10 & -0.09 & 0.06 & \textbf{0.88} & 0.03 & 0.09 & 0.33 & 0.17 &  &  & &\\
                       & -0.06 & -0.02 & -0.01 & 0.46 & 0.70 & 0.34 & -0.02 & 0.65 & 0.08 & 1.00 &  & &\\ 
                       & 0.68 & 0.68 & -0.17 & 0.56 & \textbf{0.94} & 0.01 & \textbf{0.83} & \textbf{0.97} & 0.02 &  &  & &\\
  \hline
  \multirow{3}{*}{RFC} & 0.56 & 0.30 & -0.05 & -0.05 & 0.44 & -0.06 & 0.76 & 0.03 & -0.06 & 0.60 &  & &\\
                       & 0.59 & 0.52 & 0.21 & 0.15 & 0.26 & 0.14 & \textbf{0.83} & 0.22 & -0.09 & 0.41 & 1.00 & &\\
                       & \textbf{0.83} & \textbf{0.82} & -0.07 & 0.49 & \textbf{0.86} & -0.02 & \textbf{0.96} & \textbf{0.92} & -0.02 & \textbf{0.93} &  & &\\
  \hline
  \multirow{3}{*}{TCC} & -0.07 & -0.07 & -0.03 & 0.21 & -0.04 & 0.08 & -0.04 & 0.15 & -0.02 & 0.08 & 0.02 &  &\\
                       & -0.08 & -0.06 & 0.06 & 0.33 & -0.13 & 0.44 & -0.02 & 0.31 & -0.05 & 0.26 & 0.07 & 1.00 &\\
                       & -0.14 & -0.15 & -0.06 & -0.09 & -0.08 & 0.10 & -0.07 & -0.09 & -0.08 & -0.10 & -0.09 &  &\\
  \hline                    
  \multirow{3}{*}{WMC} & 0.28 & 0.07 & -0.19 & -0.07 & 0.52 & -0.09 & 0.40 & 0.08 & 0.00 & 0.67 & 0.71 & 0.03 &\\
                       & 0.27 & 0.21 & -0.10 & 0.18 & 0.32 & 0.08 & 0.39 & 0.30 & -0.08 & 0.55 & 0.63 & 0.13 & 1.00 \\
                       & 0.69 & 0.69 & -0.14 & 0.49 & \textbf{0.92} & -0.05 & \textbf{0.87} & \textbf{0.93} & -0.01 & \textbf{0.96} & \textbf{0.94} & -0.08 &\\
\hline
\end{tabular}
\egroup
\end{table*}

\begin{table*}[h]
\caption{Metric dependencies in TuxGuitar (top row - below Q1, middle row - inter-quartile range, bottom row - above Q3) }
\label{tab:corr_tuxguitar} \centering
\bgroup
\scriptsize
\begin{tabular}{|c|c|c|c|c|c|c|c|c|c|c|c|c|c|}
\hline
Metric & CBO & DAC & DIT & ILCOM & LCOM & LD & MPC & NAM & NOC & NOM & RFC & TCC & WMC\\
  \hline
\multirow{3}{*}{DAC} & \textbf{0.92} &  &  &  &  &  &  &  &  &  &  & & \\
                     & \textbf{0.92} & 1.00 &  &  &  &  &  &  &  &  &  & & \\
                     & \textbf{0.98} &  &  &  &  &  &  &  &  &  &  & & \\
\hline
\multirow{3}{*}{DIT} & 0.64 & 0.60 &  &  &  &  &  &  &  &  &  & & \\
                     & 0.60 & 0.53 & 1.00 &  &  &  &  &  &  &  &  & & \\
                     & 0.10 & 0.05 &  &  &  &  &  &  &  &  &  & & \\
\hline
\multirow{3}{*}{ILCOM} & -0.26 & -0.23 & -0.32 &  &  &  &  &  &  &  &  & & \\
                       & -0.39 & -0.32 & -0.43 & 1.00 &  &  &  &  &  &  &  & & \\
                       & 0.08 & 0.08 & -0.15 &  &  &  &  &  &  &  &  & & \\
\hline
\multirow{3}{*}{LCOM} & -0.07 & -0.01 & -0.12 & -0.12 &  &  &  &  &  &  &  & & \\
                      & -0.17 & -0.15 & -0.19 & 0.49 & 1.00 &  &  &  &  &  &  & & \\
                      & 0.12 & 0.11 & -0.14 & 0.41 &  &  &  &  &  &  &  & & \\
\hline
\multirow{3}{*}{LD} & -0.07 & -0.04 & -0.24 & 0.41 & -0.11 &  &  &  &  &  &  & & \\
                    & -0.16 & -0.07 & -0.22 & 0.36 & 0.15 & 1.00 &  &  &  &  &  & & \\
                    & -0.08 & -0.06 & -0.18 & 0.42 & 0.06 &  &  &  &  &  &  & & \\
\hline
\multirow{3}{*}{MPC} & \textbf{0.82} & 0.66 & 0.58 & -0.25 & -0.21 & -0.02 &  &  &  &  &  & & \\
                     & 0.77 & 0.60 & 0.44 & -0.35 & -0.14 & -0.18 & 1.00 &  &  &  &  & & \\
                     & 0.49 & 0.43 & 0.09 & 0.17 & 0.54 & -0.13 &  &  &  &  &  & & \\
\hline
\multirow{3}{*}{NAM} & -0.10 & -0.05 & -0.20 & 0.30 & 0.66 & 0.09 & -0.20 &  &  &  &  & & \\
                     & -0.19 & -0.11 & -0.30 & 0.60 & 0.52 & 0.30 & -0.21 & 1.00 &  &  &  & & \\
                     & 0.10 & 0.08 & -0.18 & 0.51 & \textbf{0.84} & 0.15 & 0.52 &  &  &  &  & & \\
\hline
\multirow{3}{*}{NOC} & -0.16 & -0.13 & -0.16 & -0.10 & 0.11 & -0.07 & -0.18 & 0.04 &  &  &  & & \\
                     & 0.01 & -0.02 & -0.06 & 0.07 & 0.31 & 0.05 & 0.01 & 0.13 & 1.00 &  &  & & \\
                     & -0.06 & -0.05 & -0.08 & 0.27 & 0.12 & 0.20 & -0.03 & 0.15 &  &  &  & & \\
\hline
\multirow{3}{*}{NOM} & -0.05 & 0.03 & -0.17 & 0.12 & \textbf{0.90} & -0.03 & -0.22 & 0.75 & 0.13 &  &  & & \\
                     & -0.23 & -0.17 & -0.34 & 0.64 & 0.77 & 0.33 & -0.15 & 0.70 & 0.21 & 1.00 &  & & \\ 
                     & 0.16 & 0.14 & -0.18 & 0.50 & \textbf{0.88} & 0.14 & 0.61 & \textbf{0.92} & 0.15 &  &  & & \\
\hline
\multirow{3}{*}{RFC} & 0.79 & 0.66 & 0.47 & -0.21 & 0.26 & -0.10 & \textbf{0.82} & 0.18 & -0.10 & 0.30 &  & & \\ 
                     & 0.63 & 0.46 & 0.28 & -0.03 & 0.26 & -0.05 & \textbf{0.83} & 0.10 & 0.13 & 0.29 & 1.00 & & \\
                     & 0.35 & 0.30 & 0.06 & 0.28 & 0.61 & -0.06 & \textbf{0.90} & 0.65 & 0.03 & 0.77 &  & & \\
\hline
\multirow{3}{*}{TCC} & -0.14 & -0.08 & -0.20 & 0.24 & -0.08 & 0.48 & -0.12 & 0.12 & -0.06 & 0.11 & -0.10 & &  \\
                     & 0.05 & 0.13 & -0.05 & 0.05 & -0.18 & 0.32 & -0.03 & 0.19 & -0.05 & 0.13 & -0.02 & 1.00 & \\
                     & 0.02 & 0.02 & 0.02 & -0.14 & -0.15 & 0.02 & -0.11 & -0.11 & -0.09 & -0.18 & -0.14 & &  \\
\hline
\multirow{3}{*}{WMC} & 0.04 & 0.10 & -0.13 & 0.09 & \textbf{0.86} & 0.00 & -0.12 & 0.69 & 0.10 & \textbf{0.95} & 0.38 & 0.14 & \\
                     & -0.12 & -0.07 & -0.34 & 0.45 & 0.59 & 0.32 & 0.04 & 0.54 & 0.17 & \textbf{0.83} & 0.40 & 0.13 & 1.00 \\
                     & 0.23 & 0.21 & -0.10 & 0.29 & 0.72 & -0.06 & 0.79 & 0.72 & 0.00 & \textbf{0.84} & \textbf{0.85} & -0.17 & \\
\hline
\end{tabular}
\egroup
\end{table*}

\subsection{Longitudinal Evaluation}
\label{sec:long}
This section is dedicated to an examination of the changes to metric values during application development. Data points illustrated in Figures \ref{fig:dscode} and \ref{fig:dsdoc} are available for every metric and application version on our website. We found that values follow the illustrated distributions across all target application versions. As detailed in Section \ref{sec:qs}, maximum data points represent outliers, while minimal data points coincide with metric minimum values and are not interesting. As such, the present section is focused on discussing mean and median metric values. For the sake of brevity, we do not include all 8,560 data points. Our principle findings are that early application versions show more variability in metric values and that key application versions can be identified during which large changes to metric values occur.

\subsubsection{Metric variability in early and mature versions.} We examined the changes to metric values that occurred between consecutive versions of the same application. For all three target applications, we found that some of the most consistent changes occurred within early releases of the application. Of course, there exists no structured definition for an \textit{"early version"}, especially not one that can be used across several applications. As such, we used our familiarity with the studied applications to identify the earliest version that we considered \textit{mature}. In the case of our target applications, they were FreeMind 1.0.0Alpha4, jEdit 4.0pre4 and TuxGuitar 1.0rc1. These versions include most of the functionalities available in the latest version of the respective application, have the same look \& feel as all subsequent versions and appear to be stable software releases. Table \ref{tab:appmeans} illustrates minimum and maximum mean metric values in both early and \textit{mature} application versions.

\begin{table*}[t]
\caption{Extreme values for metric means for early (left) and \textit{mature} application versions (right). Includes data from \cite{60}}\label{tab:appmeans} \centering
\scriptsize
\begin{tabular}{c c c c c|c c c c|c c c c|}
\def\arraystretch{0.9}
  & \multicolumn{4}{c}{FreeMind} & \multicolumn{4}{c}{jEdit} & \multicolumn{4}{c}{TuxGuitar}  \\
  \cline{2-13} & \multicolumn{2}{|c}{$<$1.0.0Alpha4} & \multicolumn{2}{c|}{$\geq$1.0.0Alpha4} & \multicolumn{2}{c}{$<$4.0pre4} & \multicolumn{2}{c|}{$\geq$4.0pre4} & \multicolumn{2}{c}{$<$1.0rc1} & \multicolumn{2}{c|}{$\geq$1.0rc1} \\
  \hline
  \multicolumn{1}{|c|}{Metric} & Min & Max & Min & Max & Min & Max & Min & Max & Min & Max & Min & Max \\  
  \hline
  \multicolumn{1}{|c|}{CBO} & 3.89 & 6.15 & 5.33 & 5.57 & 3.85 & 4.29 & 4.29 & 4.91 & 6.03 & 7.56 & 7.06 & 7.88 \\
  \multicolumn{1}{|c|}{DAC} & 2.67 & 5.30 & 4.20 & 4.38 & 3.45 & 3.83 & 3.77 & 4.30 & 4.76 & 5.46 & 5.16 & 6.97 \\
  \multicolumn{1}{|c|}{DIT} & 0.15 & 1.69 & 0.70 & 1.03 & 0.37 & 0.70 & 0.32 & 0.43 & 0.45 & 0.55 & 0.78 & 1.07 \\
  \multicolumn{1}{|c|}{ILCOM} & 0.81 & 1.04 & 0.99 & 1.04 & 0.49 & 0.79 & 0.79 & 0.83 & 1.07 & 1.33 & 1.15 & 1.46 \\
  \multicolumn{1}{|c|}{LCOM} & 84.85 & 193.25 & 196.85 & 237.90 & 43.44 & 117.75 & 126.79 & 149.31 & 90.94 & 130.49 & 117.15 & 176.79 \\
  \multicolumn{1}{|c|}{LD} & 0.30 & 0.52 & 0.48 & 0.51 & 0.23 & 0.36 & 0.34 & 0.37 & 0.39 & 0.48 & 0.35 & 0.50 \\
  \multicolumn{1}{|c|}{LEN} & 11.77 & 17.07 & 16.67 & 17.17 & 12.25 & 13.10 & 13.01 & 14.35 & 14.84 & 15.09 & 15.19 & 18.26  \\
  \multicolumn{1}{|c|}{LOC} & 63.35 & 157.84 & 100.05 & 110.79 & 91.29 & 153.94 & 158.64 & 177.37 & 94.93 & 116.69 & 73.13 & 115.25 \\
  \multicolumn{1}{|c|}{LOD} & 0.72 & 0.91 & 0.78 & 0.81 & 0.73 & 0.82 & 0.73 & 0.80 & 0.68 & 0.83 & 0.88 & 0.99  \\
  \multicolumn{1}{|c|}{MPC} & 6.99 & 13.20 & 10.59 & 10.92 & 6.79 & 9.00 & 9.34 & 10.05 & 14.26 & 21.27 & 14.65 & 22.85 \\
  \multicolumn{1}{|c|}{NAM} & 7.06 & 9.84 & 9.85 & 10.09 & 5.18 & 9.02 & 8.53 & 9.19 & 9.71 & 12.13 & 9.41 & 12.98 \\
  \multicolumn{1}{|c|}{NOC} & 0.15 & 1.44 & 0.59 & 0.63 & 0.31 & 0.65 & 0.29 & 0.38 & 0.45 & 0.52 & 0.58 & 0.92 \\
  \multicolumn{1}{|c|}{NOM} & 5.26 & 7.06 & 6.88 & 6.99 & 3.16 & 5.49 & 5.28 & 5.46 & 6.38 & 7.23 & 6.13 & 8.13 \\
  \multicolumn{1}{|c|}{RFC} & 9.74 & 15.17 & 13.49 & 13.62 & 7.91 & 10.39 & 10.39 & 11.14 & 14.81 & 19.50 & 15.80 & 22.21 \\
  \multicolumn{1}{|c|}{TCC} & 0.03 & 0.16 & 0.14 & 0.16 & 0.06 & 0.13 & 0.14 & 0.17 & 0.14 & 0.22 & 0.12 & 0.18 \\
  \multicolumn{1}{|c|}{WMC} & 8.52 & 14.41 & 12.32 & 12.55 & 8.52 & 14.13 & 13.43 & 15.05 & 12.02 & 14.53 & 10.63 & 15.38 \\
  \hline
\end{tabular}
\end{table*}

We observe that for all applications, metric variability is much higher for the earlier versions. As shown in Table \ref{tab:appevolution}, the first version of FreeMind consisted of 3,722 lines of code, fewer than the first version of TuxGuitar (11,209). In contrast, the first release of jEdit (33,768 LOC) was much more mature, and already contained the application's most important functionalities. On the other hand, once the application architecture is established and the principal functionalities set is implemented, we observe a significant reduction in the variability of metric values between versions. This is illustrated for each application, in the right-hand columns of Table \ref{tab:appmeans}. Furthermore, longitudinal examination also showed that specific trends can be identified for each application with regards to how object-oriented concepts such as coupling, inheritance and structural complexity are handled. It is our opinion that additional case studies presenting a longitudinal view are required before desirable metric ranges and most importantly, reliable metric-based characterisations can be established.

\subsubsection{Causes of large variations in metric values.}  We also observed that metric values were consistent between most consecutive version pairs of the studied applications. At the same time, we could identify version pairs where metric values were greatly disrupted. We illustrate these pairs using Table \ref{tab:metricvariation}. The table also includes information about LOC and the number of classes, in order to help understand the causes behind observed variations. For example, it is obvious that a large push in development between FreeMind 0.7.1 and 0.8.0 contributed to significant changes to metric values, as evidenced by the sharp increase in application LOC and class count. The same can be said about TuxGuitar version 1.3.0. The opposite however is true for jEdit 3.0final, as well as FreeMind 0.9.0Beta17. In these versions we observe important decreases in both LOC and class count, most likely a result due to refactoring. 

\begin{table}[t]
    \caption{Application versions showing significant variance in metric values.}
    \label{tab:metricvariation}
    \centering
    \begin{tabular}{|l|c|c|c|}
    \hline
         Application & Version & LOC & Classes \\
         \hline
         \multirow{2}{*}{jEdit} & 2.6final & 46,671 & 453 \\
         \cline{2-4}
          & 3.0final & 40,756 & 282 \\
          \hline
          \multirow{4}{*}{FreeMind} & 0.7.1 & 18,928 & 199\\
          \cline{2-4}
          & 0.8.0 & 84,199 & 718\\
          \cline{2-4}
          & 0.8.1 & 84,089 & 718\\
          \cline{2-4}
          & 0.9.0Beta17 & 56,752 & 577\\
          \hline
          \multirow{2}{*}{TuxGuitar} & 1.2 & 77,056 & 736\\
          \cline{2-4}
          & 1.3.0 & 91,481 & 1,234\\
          \hline
    \end{tabular}
\end{table}

Table \ref{tab:appinflection} illustrates mean metric values for the highlighted application versions. For each version, we manually examined its source code in detail to identify the underlying changes leading to these variations. 

FreeMind 0.8.0 contains major changes, as already evidenced by the sharp increase in LOC and class count. It is the first version to use external libraries for XML processing and input forms. During use, it is clear that FreeMind 0.8.0 is more complex and fully-featured, with many changes that are visible at UI level, including more complex application preferences and features for mind map and node management. Its scope remains apparent at source file level, with only 21 out of the 92 source files remaining unchanged from 0.7.1. The number of source files also increased greatly in the newer version, from 92 to 469. Much of the observed discrepancy between numbers of source files, classes and LOC between the versions can be explained by the newer application including 272 classes that were generated by the JAXB libraries encoding most of the actions that can be performed using the application. These classes contributed with 49.434 lines to the inflation of LOC witnessed between the studied versions. Between version 0.8.0 and 0.8.1, no source files were added or deleted, but many of them have undergone small updates. This includes all generated code, that was regenerated for version 0.8.1. FreeMind again underwent significant changes for version 0.9.0Beta17, an evolution from 0.8.1. Out of 469 source files in version 0.8.1, only 127 can be found in the newer version, and all of them have undergone changes. Version 0.9.0Beta17 also added 230 new Java source files, covering all functionality areas. Action source files generated using JAXB in version 0.8.0 were replaced with a smaller number of hand-written classes with similar naming and functionality. This explains most of the class count and LOC difference between versions 0.8.1 and 0.9.0Beta17.

In the case of jEdit, version 3.0final was the only one where mean metric values were disrupted. A possible contributor to this is that relatively, early analyzed versions were more mature than equivalent ones from the other applications. In the case of version 3.0final, we observed that the package \textit{"org.gjt.sp.jedit.actions"}, which contained 153 event handler classes with low statement count and cyclomatic complexity was deleted. These were replaced with an XML file that provides action descriptors together with Java-like code snippets that are executed when the action is fired. Only 81 source files out of 341 remained unchanged between these versions.

In the case of TuxGuitar version 1.3.0, the \textit{"org.herac.tuxguitar.gui"} package was split into \textit{*.app}, \textit{*.editor} and \textit{*.graphics} packages. Most packages were updated or refactored. New plugins were added, existing ones have seen source code changes. Only 62 out of the 650 source code files remained unedited between these versions. Version 1.3.0 introduced 930 new source files, most of which contain code for custom application actions in the form of small classes having low complexity, skewing the mean and median metric values. 

The last observation is related to the expectation that mean metric values increase in more advanced application versions. Our data showed this to be true mostly in the case of FreeMind and jEdit, especially in the case of size metrics LOC, NAM and NOM. However, as we have shown in this section, this is alleviated by the refactorings that were carried out in some of the versions.

Our examination resulted in several conclusions. First, we observed that most of the significant metric variations occurred in early application versions. This was true both as highlighted in Table \ref{tab:appinflection}, as well as when manually identifying versions with significant metric variations. In addition, we feel that a more in-depth discussion is warranted regarding the effect that large numbers of small, relatively straightforward classes have on software quality characteristics. The importance and magnitude these classes should have when building metric-based models has yet to be clarified. In several cases, we observed Java source code being replaced with XML descriptors. This is an illustrative example of the inherent limitations of metric extraction tools and understanding of software based on metric values. 

\begin{table*}[t]
\caption{Mean metric values for given application versions}\label{tab:appinflection} \centering
\scriptsize
\begin{tabular}{c c c c c | c c | c c |}
  & \multicolumn{4}{c}{FreeMind} & \multicolumn{2}{c}{jEdit} & \multicolumn{2}{c}{TuxGuitar}  \\
  \cline{2-9} & \multicolumn{1}{|c}{0.7.1} & 0.8.0 & 0.8.1 & 0.9.0Beta17 & 2.6final & 3.0final & 1.2 & 1.3.0 \\
  \hline  
  \multicolumn{1}{|c|}{CBO} &  4.75 & 6.15 & 6.15 & 5.31 & 4.24 & 4.29 & 7.05 & 7.07 \\
  \multicolumn{1}{|c|}{DAC} & 3.10 & 5.29 & 5.29 & 4.14 & 3.73 & 3.82 & 5.22 & 6.30 \\
  \multicolumn{1}{|c|}{DIT} & 0.50 & 1.69 & 1.69 & 0.74 & 0.63 & 0.42 & 0.79 & 0.95 \\
  \multicolumn{1}{|c|}{ILCOM} & 0.95 & 0.80 & 0.80 & 1.03 & 0.52 & 0.77 & 1.43 & 1.22 \\
  \multicolumn{1}{|c|}{LCOM} & 179.54 & 152.56 & 152.56 & 189.32 & 47.95 & 114.37 & 176.79 & 130.11 \\
  \multicolumn{1}{|c|}{LD} & 0.42 & 0.43 & 0.43 & 0.51 & 0.25 & 0.36 & 0.50 & 0.35 \\
  \multicolumn{1}{|c|}{LEN} & 15.23 & 16.91 & 16.91 & 17.06 & 12.62 & 12.97 & 15.45 & 18.26 \\
  \multicolumn{1}{|c|}{LOC} & 102.95 & 157.83 & 157.52 & 97.87 & 100.16 & 151.28 & 115.24 & 80.74 \\
  \multicolumn{1}{|c|}{LOD} & 0.86 & 0.72 & 0.72 & 0.80 & 0.82 & 0.73 & 0.89 & 0.98 \\
  \multicolumn{1}{|c|}{MPC} & 11.53 & 13.19 & 13.19 & 10.51 & 7.58 & 9.00 & 22.82 & 14.64 \\
  \multicolumn{1}{|c|}{NAM} & 9.08 & 8.99 & 8.99 & 9.77 & 6.01 & 9.01 & 12.96 & 9.85 \\
  \multicolumn{1}{|c|}{NOC} & 0.37 & 1.44 & 1.44 & 0.61 & 0.56 & 0.33 & 0.65 & 0.64 \\
  \multicolumn{1}{|c|}{NOM} & 6.61 & 7.06 & 7.06 & 6.82 & 3.72 & 5.48 & 8.12 & 6.31 \\
  \multicolumn{1}{|c|}{RFC} & 13.07 & 15.16 & 15.16 & 13.29 & 8.84 & 10.38 & 22.20 & 15.80 \\
  \multicolumn{1}{|c|}{TCC} & 0.06 & 0.08 & 0.08 & 0.15 & 0.08 & 0.12 & 0.16 & 0.12 \\ 
  \multicolumn{1}{|c|}{WMC} & 12.94 & 14.40 & 14.38 & 12.16 & 9.27 & 14.00 & 15.38 & 11.30 \\
  \hline
\end{tabular}
\end{table*}

\subsection{Threats to Validity}
We carried out our study using the following steps, in order: preparing application versions, extracting metric data, processing the metric data and analysing it. We presented all the steps required to duplicate our study in detail. Extracted metric information, as well as aggregated data used for analysis is available on our website. Each target application version was manually examined in order to ensure that no factors that could influence metric values were present. We provided structured definitions for all metrics used, and extracted the data using a freely-available, cross-platform tool.

We selected three similar applications from a programming language and architecture standpoint. This helps limit external threats to validity related to application selection and generalization of results. This also allows comparing obtained results, as all three applications include the same layers. Application selection and metric extraction were finalized before data analysis, to eliminate selection bias. All results are presented both individually, per-application, as well as in aggregate form.

However, we believe one of our most important contributions was the comparative evaluation against a large-scale cross-sectional study that was carried out using the same methodology as ours. We believe this will help create a solid basis for additional studies towards a metric-based understanding of software quality and the software development process. 

Among existing threats, we must include the limited number and types of studied applications. This means that additional research is required in order to draw conclusions about other types of software, such as non GUI-driven or mobile applications. Furthermore, as we only included open-source software, they might not be representative for other applications. As such, we believe that additional experimental evaluation is required in order to cover additional applications, programming languages as well as considered metrics.

\section{Conclusions and Future Work}
In this paper we establish a number of metrics that previous research has associated with software product quality. We select three open-source, user interface-driven applications developed in Java and analyze the values and relations between these metrics within each application's entire development history.

Each step of our evaluation is detailed and we employ open-source tooling to ensure that our evaluation is repeatable. At each step, we compare our results with a comparable large-scale evaluation, obtaining results from an aggregate of over 250\footnote{\cite{24} evaluated 146 software projects} application versions. We believe these combined results provide a sound foundation to be used in further research.

We found that metric distributions, mean, median and modus values were consistent across the studies. Mean and median values prove stable once applications reach maturity, as evidenced in all three target applications. Comparing values across studied applications revealed the existence of trends in metric values, driven by the architecture and design of the underlying application.

With regards to identified metric dependencies, we could identify metric pairs showing strong correlation across applications and application versions, as well as certain metrics that did not show correlation with any others. We further investigated the confounding effect of class size in order to confirm our findings.

Our longitudinal approach also revealed that across many application version we could not witness significant changes to aggregated metric values. Where such changes occurred, they were mostly driven by application development as well as refactoring, and were reflected in object-oriented metric values. 

An important avenue for further research regards a finer grained analysis, in order to detect significant changes at package and class levels, not just those that are visible at aggregated level. Our evaluation should be extended in order to cover other application types, including mobile and non user interface-driven software. We believe this type of research can lay the foundation for identifying suitable metric thresholds that point toward good design practices. Another aspect regards the role played by the programming language itself, as it too plays an influence on metric values. 

The end goal of this research is represented by a characterization of good design and development practices, where software metrics will have an important role for understanding and controlling the software development process.

\bibliographystyle{splncs04}
\bibliography{references}

\end{document}